\documentclass[10pt,journal,compsoc]{IEEEtran}

\IEEEoverridecommandlockouts
\usepackage{cite}
\usepackage{amsmath,amssymb,amsfonts}
\usepackage{algorithmicx}
\usepackage{graphicx}[pdf]
\usepackage{textcomp}
\usepackage{xcolor}
\usepackage{subcaption}
\usepackage{algorithm}
\usepackage{algpseudocode}
\usepackage{breqn}
\usepackage{mathtools}
\usepackage{bm}
\usepackage{esvect}
\def\BibTeX{{\rm B\kern-.05em{\sc i\kern-.025em b}\kern-.08em
    T\kern-.1667em\lower.7ex\hbox{E}\kern-.125emX}}
    
\DeclareMathOperator*{\argmin}{argmin}   

\newtheorem{definition}{Definition}

\begin{document}

\title{Multi-Use Trust in Crowdsourced IoT Services}



\author{Mohammed Bahutair, Athman Bouguettaya ~\IEEEmembership{Fellow,~IEEE}, and Azadeh Ghari Neiat}

\markboth{IEEE Transactions on Services Computing}%
{Bahutair \MakeLowercase{\textit{et al.}}: Multi Usage-Based Trust in Crowdsourced IoT Services}

\IEEEtitleabstractindextext{%
\begin{abstract}
We introduce the concept of \emph{adaptive trust} in crowdsourced IoT services. It is a customized fine-grained trust tailored for specific IoT consumers. \emph{Usage patterns} of IoT consumers are exploited to provide an accurate trust value for service providers. A novel \emph{adaptive trust management framework} is proposed to assess the dynamic trust of IoT services. The framework leverages a novel detection algorithm to obtain \emph{trust indicators} that are likely to influence the trust level of a specific IoT service type. Detected trust indicators are then used to build \emph{service-to-indicator} model to evaluate a service's \emph{trust at each indicator}. Similarly, a \emph{usage-to-indicator} model is built to obtain the \emph{importance of each trust indicator} for a particular usage scenario. The per-indicator trust and the importance of each trust indicator are utilized to obtain an overall value of a given service for a specific consumer. We conduct a set of experiments on a real dataset to show the effectiveness of the proposed framework.
\end{abstract}

}

\maketitle

\IEEEdisplaynontitleabstractindextext

\IEEEpeerreviewmaketitle

\IEEEraisesectionheading{\section{Introduction}\label{sec:introduction}}

\label{introduction}

\IEEEPARstart{T}{he} rapid development in WiFi and System-on-Chip technologies paved the way for Web-enabled devices \cite{atzori2010internet}. The interconnection between such devices led to the emergence of the \emph{Internet of Things (IoT)}. More specifically, the IoT consists of all Internet-enabled \emph{things} equipped to collect data and communicate with other things. IoT has facilitated the emergence of key applications such as smart cities and and smart homes \cite{gubbi2013internet, shao2016clustering}. The concurrent emergence of crowdsourcing and service oriented computing \cite{de2011building, yu2012multi} is providing an ideal platform for the rapid and effective deployment of IoT applications whereby IoT devices crowdsource services from other IoT devices in dynamic environments \cite{ren2015exploiting}. Examples of such services include the use of processing power and WiFi access of other nearby IoT devices \cite{habak2015femto}. \emph{Crowdsourced services} are generally defined as services provided by the crowd to the crowd \cite{yuen2011survey}. \emph{IoT service crowdsourcing} is defined as the provisioning of services to other nearby IoT devices in an open and dynamic environment. We refer to such services as \emph{crowdsourced IoT services}. IoT devices can crowdsource a wide range of types of services, including \emph{computing resources} \cite{habak2015femto},  \emph{energy sharing} \cite{lakhdari2018crowdsourcing,dhungana2019exploiting}, and \emph{environmental sensing} \cite{kelly2013towards}. For example, the crowdsharing\footnote{Crowdsharing and crowdsourcing are used interchangeably in this paper.} of energy services involves wirelessely sending energy from an IoT device (service provider) to one or more other nearby IoT devices (service consumers) \cite{lakhdari2018crowdsourcing}.

\begin{figure*}
    \centering
    \begin{subfigure}[b]{0.30\textwidth}
        \includegraphics[width=\textwidth]{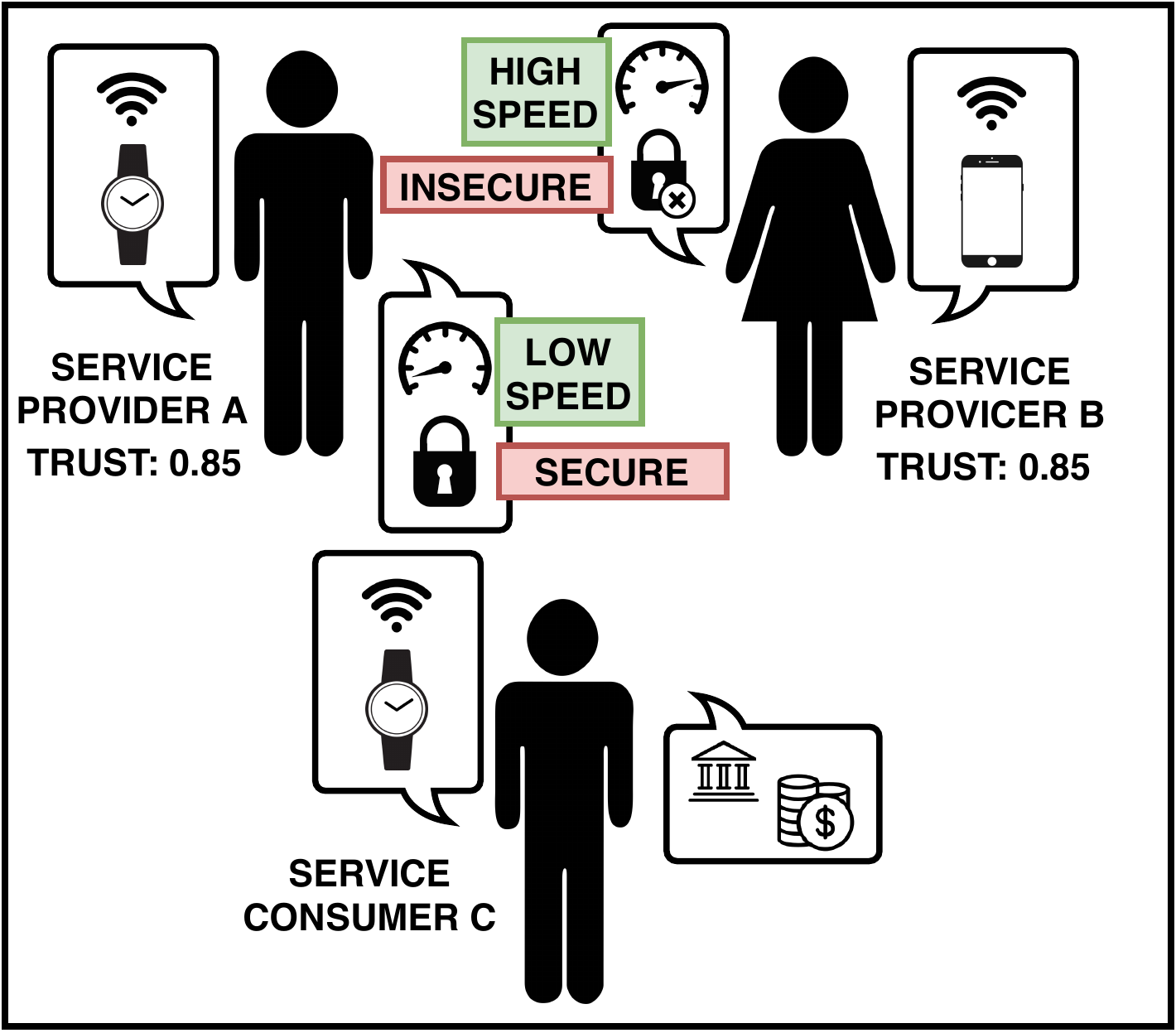}
        \caption{}
        \label{fig:scenario1}
    \end{subfigure}
    \hspace{1.0cm}
    \begin{subfigure}[b]{0.30\textwidth}
        \includegraphics[width=\textwidth]{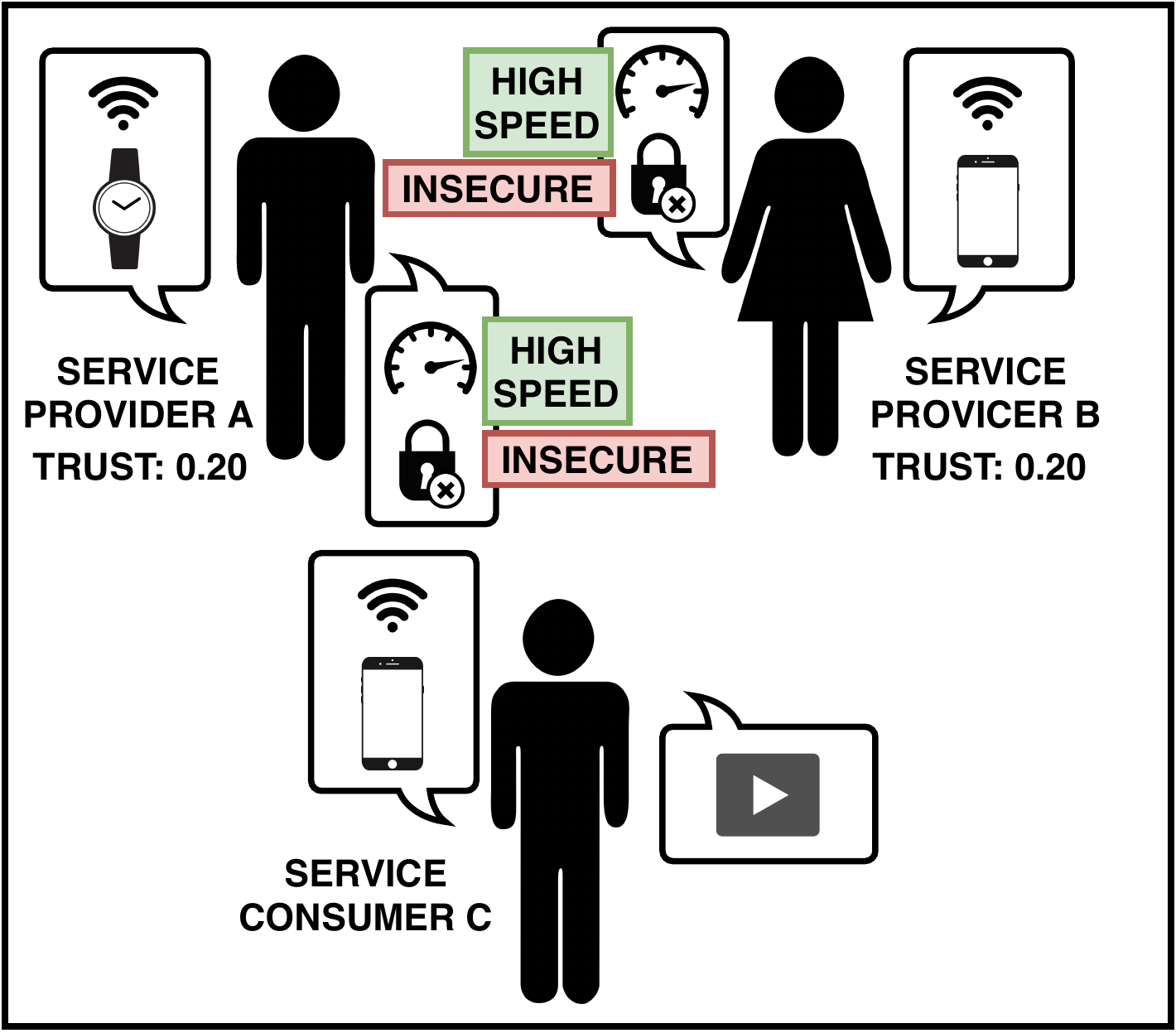}
        \caption{}
        \label{fig:scenario2}
    \end{subfigure}
    \caption{Examples of how a fixed trust value for IoT services may fail in representing the service based on the intended usage.}
    \label{fig:scenarios}
\end{figure*}

IoT crowdsourcing platforms provide a myriad of opportunities in terms of \emph{convenience} and \emph{resource utilization} \cite{habak2015femto}. However, crowdsharing services in such a highly dynamic environment as in IoT poses some fundamental challenges in ensuring the  \emph{trusted} exchange of services \cite{bahutair2020just, bahutair2019adaptive}. In particular, there is a need to provide a trusted framework that enables the free, safe, and secure exchange of IoT services where there is no central authority. The successful deployment of any IoT crowdsourcing platform is predicated on providing a trust framework for assessing and evaluating individual trust levels of each participating IoT service. It is expected that IoT service consumers would expect that consuming services would require trusting IoT service providers. For example, an IoT computing services' provider would involve providing their computation resources, e.g., CPU, to other nearby devices \cite{habak2015femto}. Consumers, however, have no guarantee that the results of their data has not been disclosed. Another example is WiFi hotspot services \cite{neiat2017crowdsourced}, where IoT providers offer to share their Internet access to nearby devices using WiFi hotspots. IoT consumers who are in need of Internet access can consume those services. However, because of the anonymity and loss of central authority, consumers cannot be sure that the service is reliable, or their data is secured. These concerns may be alleviated by assessing the trustworthiness of the providers prior to service consumption. IoT crowdsourcing environments, however, exhibit specific characteristics that make trust assessment challenging. Such characteristics include the \emph{diversity (i.e., heterogeneity)} and \emph{anonymity} of IoT users and devices, and the lack of a central managing authority.

The dynamic nature of IoT services poses a significant challenge in assessing their trust. The dynamism of such services is caused by several factors, one of which is \emph{the services' usability}. Unlike traditional services, a single IoT service can serve different \emph{uses}. For example, a WiFi hotspot service \cite{neiat2017crowdsourced} can be \emph{used} for either shopping on eBay or watching a movie on Netflix. Traditional trust management frameworks do not consider \emph{how} a service is used. They evaluate services by providing an aggregated trust value for the service \cite{saied2013trust,chen2011trm}. The obtained trust value is then used by a potential consumer to judge a service \emph{regardless of their usage}. However, a single trust value may not be sufficient to determine how trustworthy a service is, given a specific consumer usage. For example, assume an IoT WiFi service has a relatively low trust value due to its low Internet speed. The service, however, is very secure and capable of preserving its consumers' privacy. The given service can be \emph{used} differently across multiple consumers, e.g., shopping, surfing the web, making calls, etc. The service's \emph{usability} should be considered to provide an accurate assessment of the service's trust. On one hand, a consumer might use the service for initiating bank transactions. The \emph{security} of the service is more crucial than its Internet speed. The given service, therefore, should have a higher trust value since it delivers a highly secured Internet success. On the other hand, a consumer might use the service for streaming videos. In such a case, the service's trust should be lowered as streaming videos require high Internet speed, which is not delivered by the service.

We propose to address the challenges above using an \emph{adaptive trust framework} that evaluates the trustworthiness of a specific IoT service based on consumers' \emph{usage patterns}. The proposed framework will enable a service to be evaluated differently according to the usage of the consumer. Additionally, a single consumer may use a service for different uses simultaneously (e.g., using a WiFi hotspot service for shopping on Amazon and listening to music). In such a case, the framework should be able to evaluate a given service based on \emph{multiple uses} by a single consumer. Therefore, we propose an approach that considers the case of \emph{multiple uses} when assessing the trust level of IoT services. More formally, the contribution of our work is as follows:
\begin{itemize}
    \item A trust management framework that evaluates the \emph{adaptive trust} for IoT services based on consumers' \emph{usage}. The proposed framework consists of four stages. The first stage detects the \emph{indicators} that may govern the trustworthiness of a service for a specific \emph{service type}. The second stage builds a \emph{service-to-indicator model} that evaluates a service's trustworthiness at each detected trust indicator. The third stage creates a \emph{usage-to-indicator model} to obtain the expected trust level of each indicator for a given \emph{usage}. The final stage leverages the results from the two previous stages to provide a trust value that \emph{adapts} to the given usage.

    \item A two-stage approach that accounts for \emph{multi-usage scenarios}. The first stage predicts the usage pattern of a given IoT service consumer. The prediction helps to generate a set of possible uses by a specific consumer ahead of time. The second stage computes the expected trustworthiness of each usage and \emph{aggregates} them into a single trust expectation. The aggregated trust expectation provides a general representation for the usage pattern.
    
\end{itemize}

The rest of the paper is organized as follows. Section \ref{sec:pre} presents essential preliminaries and defines the problem. Sections \ref{section:framework} and \ref{section:multi_usage} introduce our adaptive trust framework. Section \ref{sec:evaluation} discusses the results of our experiments. Section \ref{sec:conclusion} concludes our paper.

\subsubsection*{Motivation Scenario}
We use two use-case scenarios to illustrate the importance of our work. Assume an IoT crowdsourcing platform where IoT devices provide and consume WiFi hotspot services between each other. WiFi hotspot services can be provided using WiFi sharing apps, such as Open Garden\footnote{https://opengarden.com}. Users of the crowdsourcing platform can use their IoT devices (e.g., smartphone, smartwatches, etc) to provide/consumer services. Assume three different users $A$, $B$, and $C$. Users $A$ and $B$ provide WiFi hotspot services through their smartphone and smartwatch, respectively (service providers). User $C$ wishes to consume a WiFi hotspot service using their smartphone (service consumer). Any given service should have its trustworthiness evaluated before it is consumed. Finally, we assume that the trustworthiness of a service is governed by two \emph{trust indicators}: \emph{security} and \emph{Internet speed}.

In the first scenario, we assume that both providers $A$ and $B$ have equal trust values of 0.85, i.e., high trust level (trust values are assumed to be acquired using aggregated consumer ratings or traditional trust frameworks). Assume consumer $C$ wants to use a WiFi hotspot service for performing bank transactions. In this scenario, consumer $C$ is presented with two providers that have the same trust level. Providers $A$ and $B$, however, obtain their high trust level for two different reasons. On one hand, provider $A$ has a highly secured Internet and relatively low Internet speed. On the other hand, provider $B$ offers a high Internet speed with relatively lower security. Consumer $C$'s usage requires a highly secured service since sensitive information is to be shared. As a result, provider $A$ should be selected despite both providers have the same trust values. Therefore, the trustworthiness of providers $A$ and $B$ should be \emph{adjusted}, i.e., a higher value for provider $A$ and lower value for provider $B$.

In the second scenario, providers $A$ and $B$ have a low trust value, i.e., 0.2 (trust values are assumed to be acquired using aggregated consumer ratings or traditional trust frameworks), see Fig. \ref{fig:scenario2}. We assume that the low trust value is due to providing insecure WiFi hotspot services. The two services, however, provide high Internet speed. In other words, the low trust value is the result of having a low-security level, not the service's speed. Assume that consumer $C$ wishes to use a WiFi hotspot service to stream publicly available videos. As stated earlier, consumers need to assess providers before using their services. In this scenario, consumer $C$ is presented with two providers that have low trust values due to their low security. Consumer $C$'s usage, however, favors Internet speed over its security, i.e., playing videos requires high Internet speed for reliable playback. Both providers offer high Internet speed. Therefore, the trustworthiness of providers $A$ and $B$ should be \emph{adjusted} for consumer $C$'s usage, i.e., their trust value should be increased.

It is worth noting that adapting to consumers' usage while assessing trust is not specific to a single service type, e.g., WiFi hotspot services. For example, assume a crowdsourcing platform where IoT devices share their computational resources (e.g., memory and processors) to other IoT devices. In such a case, IoT devices with limited resources (service consumers) can send tasks to other computationally-capable devices (service providers) for processing. A consumer with a smartwatch can ask a service provider to render Google Map images in real-time. Other consumers might wish to perform computational tasks on a confidential Excel file. On one hand, the map rendering scenario may require more processing power. On the other hand, sending a confidential file to a service requires privacy assurances.

\begin{figure}
    \centering
    \includegraphics[width=0.35\textwidth]{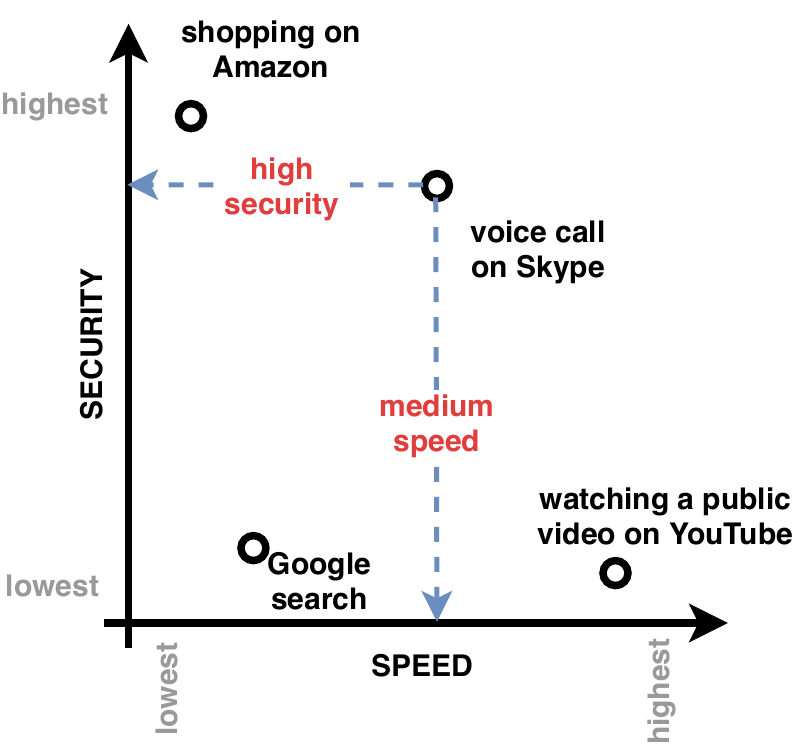}
    \caption{Example of four different uses and their expectation from a WiFi hotspot service.}
    \label{fig:two_dimensional_trust}
\end{figure}

\begin{figure}
    \centering
    \includegraphics[width=0.48\textwidth]{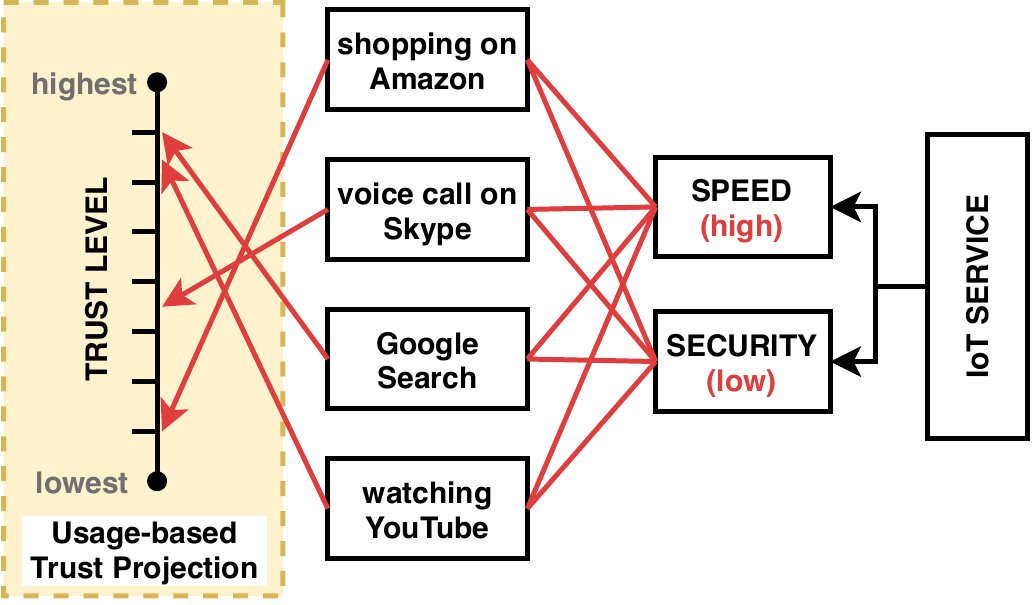}
    \caption{Example of different service trust values for different uses.}
    \label{fig:trust_usage_projection}
\end{figure}

\section{Preliminaries and Problem Definition}
\label{sec:pre}

As stated earlier, the trust level of an IoT service may be influenced by how it is used. We use the following example to illustrate the importance of considering service usage in trust assessment. Assume an IoT device (e.g., smartphone), that provides Internet access using WiFi hotspot (the IoT service). Additionally, assume four potential consumers. Each consumer wishes to consume the service for one of the following \emph{uses}: 1) shopping on Amazon, 2) watching a publicly available video on YouTube, 3) voice calling using Skype, and 4) searching on Google. Each usage has a set of expectations from the service. The service is considered trustworthy if it delivers on those expectations. A single expectation falls under one ``category", e.g., watching a video on YouTube expects high (expectation) Internet speed (category). For simplicity, we consider two categories in this example; Internet Speed, and Security. Fig. \ref{fig:two_dimensional_trust} shows the expectations for each of the four uses. There are two ``dimensions" since two categories are considered. Each dimension represents the expectation of a usage at a certain category. For instance, voice calling on Skype may require a highly secure Internet access since private data may be transferred during the call. In terms of speed, a Skype voice may expect normal Internet speed to operate reliably. Each usage can have different expectations from the IoT service as shown in Fig. \ref{fig:two_dimensional_trust}. Therefore, it is crucial to reflect these variations on the service's trustworthiness. In other words, the trust level of the service should ``adapt" based on how the service is used. Assume that the IoT service provides high Internet speed and insecure Internet access. Fig. \ref{fig:trust_usage_projection} shows how the service's trust level changes by varying the usage. One way to represent trust evaluation for a given service is to ``project" the service on the usage we are interested in. The result of this projection is the service's trust level from the perspective of that usage. In Fig. \ref{fig:trust_usage_projection}, the service is projected on each of the four potential uses. The expectation of the uses is utilized to generate the output of the projection; i.e., a trust value "adapted" to the usage. Recall that each usage can have distinct expectations. As a result, the service's trustworthiness should also vary to adapt to those different expectations.

\subsection{Trust and Security}
\emph{Trust} and \emph{security} have been used interchangeably in the domain of networking and distributed system applications \cite{yan2014survey}. \emph{Security may be one of several options to establish trust} in crowdsourced IoT environments. On one hand, security is concerned with preserving the confidentiality and integrity between two or more parties. Encryption methods are usually the means to implement security goals. For example, symmetric and asymmetric encryption algorithms (e.g., AES \cite{heron2009advanced} and RSA \cite{rivest1978method}) can be used to prevent unauthorized access to data (preserving confidentiality). As for integrity, blockchain \cite{nakamoto2019bitcoin} is one of the recent technologies that can be used to protect data from unwanted manipulation or alteration without the involvement of a trusted third party. On the other hand, \emph{trust is at a higher level of abstraction than security}, as it could be implemented and delivered using a range of solutions including but not limited to security. Trust is defined as \emph{the probability that an entity meets an expectation} \cite{mcknight2000trust, yan2014survey, sherchan2013survey, chen2011trm}. For example, assume an IoT crowdsourcing environment where devices \emph{can wirelessly share energy} among each other \cite{lakhdari2018crowdsourcing}. A service provider may advertise their ability to provide 800mW within 30 minutes. Therefore, consumers \emph{expect} that they would get what has been advertised. In this example, security does not guarantee the delivery of the expectation. There can be several other goals to trust that cannot be fulfilled simply by only guaranteeing security. Such goals include but are not limited to \emph{reliability}, \emph{availability}, and \emph{ability} \cite{yan2014survey}. To illustrate this, assume a crowdsourcing scenario, where IoT devices use an IoT device as a WiFi hotspot service. Given a provider and consumer, security may be achieved by encrypting the communication channel, essentially preventing others from intercepting the channel. This, however, does not guarantee the trustworthiness of the provider or consumer. The provider might provide unreliable Internet (poor Internet speed, hence untrusted). Additionally, security measures such as encryption can protect both providers from external attacks but may fail against internal ones \cite{saied2013trust}. Therefore, the consumer (or provider) may succeed in sending malware to the other although the channel is encrypted. 

\subsection{Problem Definition}
\begin{definition}
We represent a given IoT crowdsourced service $S$ by the following tuple: $<id, o, d, f, q>$ where,
\end{definition}
\begin{itemize}
    \item $id$ is a unique service ID. 

    
    \item $o$ is the IoT device's owner. The IoT device's owner can be an individual, business, or cooperation (e.g., restaurants or universities). We assume in this paper the owner of the device is also the service provider (i.e., $p_S = o$).
    
    \item $d$ is the IoT device that offers the service $S$ (e.g., smartwatch).
    
    \item $f$ is a set of functions offered by $S$ (e.g., computing resources). 
    
    \item $q$ is a set of all non-functional (QoS) parameters of  $S$ (e.g., CPU cores and memory capacity). 
\end{itemize}


\begin{definition}
\emph{service attributes} are defined as the set of \emph{inherent characteristics} of an IoT service. Such characteristics can be leveraged to uniquely identify a given IoT service from other services. We denote the set of service attributes as $A_S$, where $a \in A_S$ is a single service attribute. The set $A_S$ is a hyperparameter and assumed to be picked by an expert.
\end{definition}
Examples of service attributes are: the device's operating system, the owner's reputation, the device being used to provision the service, etc.

In a crowdsourced IoT environment, IoT service consumers search for available nearby service providers. Once a given consumer $c_S$ finds a service provider $p_S$, the consumer starts to consume their service. The services typically have a single functionality, e.g., providing Internet access via WiFi hotspot. \emph{Consumers, however, can use a given service differently despite the service having one functionality.}

\begin{definition}
We define the \emph{service usage} $u$ as how a given IoT service consumer uses a particular service $S$.
\end{definition}

For instance, the Internet provided by a WiFi hotspot service can be used to either shop online (a service usage), or watch videos (another usage). We assume that a given service usage $u$ is associated with a set of \emph{metadata} $M_{u}$. A single metadata $m \in M_{u}$ is a word that describes an aspect of the usage $u$. For example, `Skype calling' is potential usage for WiFi hotspot services. `Streamed', `Personal', and `Prolonged' are all relevant metadata that describe the usage `Skype calling'. A specific consumer can use the service for \emph{multiple purposes} during a single service session. For example, a consumer can message their friends, browse the Web, and listen to music using a WiFi hotspot service. We refer to the set of possible uses during a service session as the \emph{consumer's usage pattern} $\mathcal{U}_c$.

As stated earlier, a given service can be used differently despite having a single functionality. Different consumers may assess a particular service's trust differently based on their usage patterns $\mathcal{U}_c$. It is, therefore, crucial to account for the service usage when evaluating the trustworthiness of IoT services. More formally, we define \emph{\textbf{adaptive trust $\mathcal{T}_{Adapt}$} as the customized trust for a given usage pattern}. The purpose of our work is to identify a function $\mathcal{F}(S, \mathcal{U}_c)$ such that:
\begin{equation}
    \mathcal{T}_{Adapt} \approx \mathcal{F}(S, \mathcal{U}_c)
\end{equation}

\section{Adaptive Trust Assessment Framework} 
\label{section:framework}

\begin{figure*}
    \centering
    \includegraphics[width=0.9\textwidth]{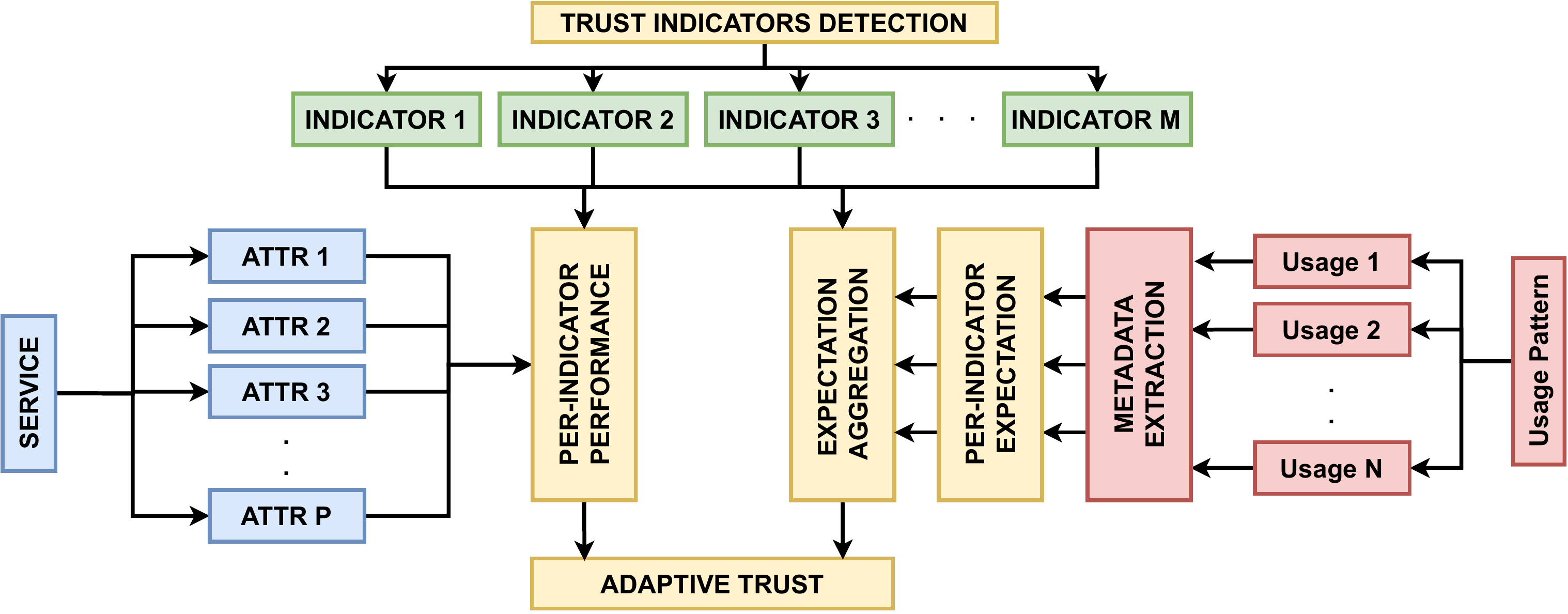}
    \caption{Framework overview}
    \label{fig:framework_overview}
\end{figure*}
We present a trust management framework that assesses the trustworthiness of IoT services (Fig. \ref{fig:framework_overview}). The proposed framework evaluates a given service's trust based on the \emph{usage patterns} of its consumers. As a result, each consumer would have a trust assessment that is specific only to their usage pattern. In other words, the trust value of a service would \emph{adapt} based on how it is used. The framework consists of four stages: \emph{trust indicators detection}, \emph{per-indicator trustworthiness prediction}, \emph{per-indicator expectation}, and \emph{adaptive trust assessment}.

For any given crowdsourced IoT service type, e.g., crowdsourced WiFi hotspots, there can be a set of \emph{indicators} that influence the trust level of a service. For example, indicators that may impact WiFi hotspot services' trust include Internet speed, connection security, and availability. Such indicators are not fixed and can vary across different IoT service types (e.g., WiFi hotspot, computing resources, sensing, etc.). Therefore, the first stage of the proposed framework (i.e., \emph{trust indicators detection} stage) aims at identifying service type-specific \emph{indicators} that may influence the trustworthiness of a service. The \emph{per-indicator trustworthiness prediction} stage computes how well a service performs at each detected indicator. The \emph{per-indicator expectation} stage evaluates the importance of each indicator given a specific usage pattern. Finally, a usage-based adapted trust is computed at the \emph{adaptive trust assessment} stage using the results obtained from the last two stages.

It is worth noting that the framework's stages do not execute sequentially. The execution flow of the framework is split into two phases: \emph{online} and \emph{offline} phases. We refer to the online phase as the service provisioning/consumption phase, i.e., the time when IoT devices provide and consume services. Any time before service provisioning/consumption is referred to as the offline phase. The \emph{trust indicators detection} stage is executed entirely during the offline phase. The \emph{per-indicator trustworthiness prediction} and \emph{per-indicator expectation} stages occur in the online and offline phases (see Sections \ref{section:performance_prediction} and \ref{section:expectation} for further details). The \emph{adaptive trust assessment} stage executes entirely in the online phase. 

\subsection{Trust Indicators Detection}
\label{section:indicators_detection}
IoT devices can provide and consume services of different types, e.g., WiFi hotspots \cite{neiat2017crowdsourced}, computing \cite{habak2015femto}, sensing \cite{ganti2011mobile}, green energy \cite{lakhdari2018crowdsourcing}, etc. The attributes that control services' trust can vary across different service types. For example, the trust of computing services \cite{habak2015femto} may be affected by the following attributes: processing power, memory capacity, and data privacy. However, the trustworthiness of sensing services \cite{ganti2011mobile} may be governed by another set of attributes including freedom of movement, sensors' efficiency, and communication speed. We refer to such attributes as \emph{trust indicators} $\mathcal{F}$, since they provide an \emph{indication} of how trustworthy a service can be. The proposed framework aims at identifying trust indicators that are crucial for a given consumer usage pattern. A trust value is then computed which is tailored to that usage pattern.

We leverage rating data to identify the number of trust indicators. We assume that the data has been previously collected during previous interactions between service providers and consumers. We assume that the data is maintained in a decentralized and integrity preserving platform (e.g., \cite{bahutair2021blockchain}). In such interactions, consumers gave ratings to services after consumption. All previously provided/consumed services are assumed to be of the same service type (e.g., WiFi hotspot services). The data includes the \emph{usage} of consumers. A sample in the data is represented by the tuple: $<s, u, r>$, where $s$, $u$, and $r$ refer to the service, usage, and given rate, respectively. \emph{The anomalies found between $r$ values for a specific service $s$ is used to infer the number of trust indicators}. For instance, assume a service $A$ in the rating data that has been rated 10 times, of which five are 9/10 and the rest is 3/10. We can infer that service $A$ performed well at certain aspects (trust indicators) evident by the high ratings (i.e., 9/10) for some uses. Additionally, service $A$ behaved poorly in some other aspects since half of the ratings are low. \emph{The correlation between different uses across different service instances} can also be used to identify the number of potential trust indicators. For example, assume two consumers with uses $X$ and $Y$ both rated services $A$, $B$, and $C$ with 9/10, 2/10, and 5/10, respectively. We can conclude that uses $X$ and $Y$ may depend on similar trust indicators since both of them have correlated rates. That is, their ratings are similar across different services. Algorithm \ref{alg:detect_indicator_number} lists the steps to detect the number of indicators for a specific IoT service type.

\begin{algorithm}
    \renewcommand{\algorithmicrequire}{\textbf{Input:}}
    \renewcommand{\algorithmicensure}{\textbf{Output:}}
    \caption{Identifying the Number of Trust Indicators}
    
    \label{alg:detect_indicator_number}
    \begin{algorithmic}[1]
        \Require
        $D$: rating data for a particular IoT service type,
        \Ensure $|\mathcal{F}|$: the number of trust indicators
        
        \State $D_{\text{grouped}} = D$.groupBy($service\_id$)
        \State $C_0$ = $D_{\text{grouped}_0}$
        \For{$C_i = D_{\text{grouped}_i}, i=1 \to |D_{\text{grouped}}|$}
            \For{$c \in C_{i-1}$}
                \State $C_{sim}$ = Find clusters similar to $c$ in $C_i$
                \If{$|C_{sim}|$ == 1}
                    \State $c = c \cup C_{sim}$
                \Else
                    \State Add all elements in $c$ that do not exist in  $C_{sim}$ to all clusters in $C_{sim}$
                    \State $C_{sim}$.addAll($C_{i-1}$)
                    \State $C_{i-1}$.remove($c$)
                \EndIf
                \State $C_i = C_{i-1}$
            \EndFor
        \EndFor
        \State $|\mathcal{F}| = |C_{last}|$
        \State \Return $|\mathcal{F}|$
    \end{algorithmic}
\end{algorithm}

The input to Algorithm \ref{alg:detect_indicator_number} is previously discussed rating data, whereas the output is the number of trust indicators for a particular IoT service type, e.g., sensing services. The first part tries to identify the rating anomalies for a specific service. Therefore, we group rating samples by services (line 1) and cluster the samples in each groups based on rating values $r$ (line 4). Lines 5 - 14 compare between every two groups to find correlated samples. For each cluster in the first group, Line 6 finds all clusters in the second group that share some elements with the former cluster. If one cluster is found (Line 7-8), the two clusters are merged together. Otherwise, the cluster is substituted with the discovered clusters (Lines 10 - 15).  


\subsection{Per-indicator Trustworthiness Prediction}
\label{section:performance_prediction}

\begin{figure}
    \centering
    \includegraphics[width=0.40\textwidth]{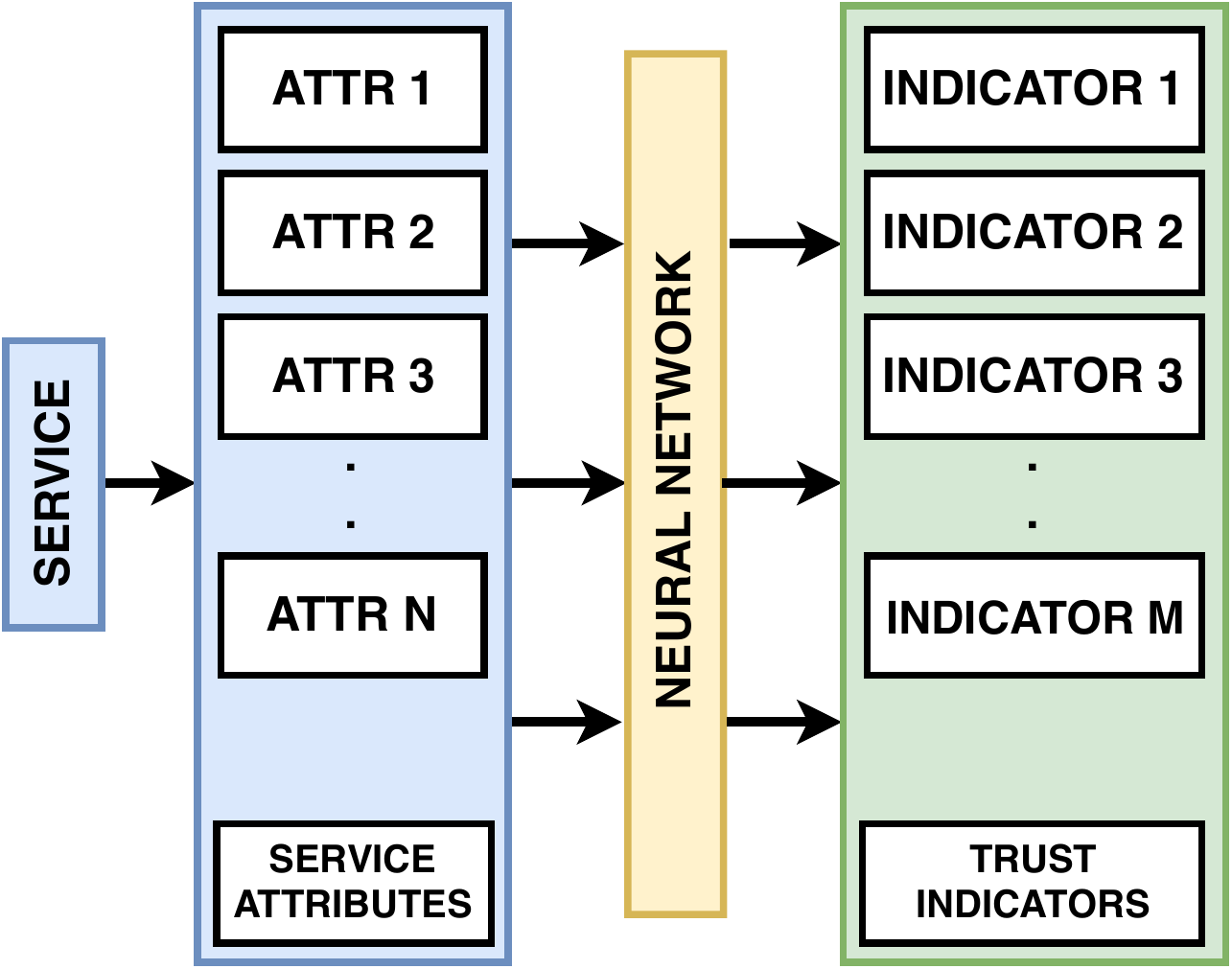}
    \caption{Service-to-Indicator Model}
    \label{fig:service_to_indicator_model}
\end{figure}
An IoT consumer evaluates the trustworthiness of a service provider before consuming their services. As stated previously, several trust indicators may influence the trust level of IoT services. The goal of the per-indicator trustworthiness prediction stage is to assess IoT services' trustworthiness at each given indicator. The execution of the stage is split between the offline and online phases. Recall that the offline phase happens before service provision/consumption while the online phase occurs when services are being provided and consumed. During the offline phase, a \emph{service-to-indicator} model $\left(\mathcal{M}_{\mathcal{F}}^{\mathcal{S}}\right)$ is trained (we assume that the model is trained on edge servers). The model is later used by IoT consumers to evaluate the trust level of IoT service providers during the online phase. The model predicts the trust level of a given IoT service at each available trust indicator. We use Neural Networks (NN) \cite{haykin2004comprehensive} to train and build the service-to-indicator model $\left(\mathcal{M}_{\mathcal{F}}^{\mathcal{S}}\right)$, see Fig. \ref{fig:service_to_indicator_model}. The use of Neural Networks stems from the following two reasons: (1) Neural Networks have the ability to identify complex non-linear relationships between input and output variables \cite{tu1996advantages} (2) the trustworthiness at each indicator can be extracted easily from the output layer.

Neural Networks consist of a set of interconnected nodes called \emph{neurons} that are organized as \emph{layers}. Links that connect nodes to each other are called \emph{synapses}. The first layer in the network is referred to as the \emph{input layer}, whereas the last layer is referred to as the \emph{output layer}. Layers between the input and output are called \emph{hidden layers}. Every synapse in the network has a weight and each neuron has value called the \emph{activation}. For a given neuron, its activation value is computed using a function that takes as parameters the weights of all incoming links. Such functions are referred to as \emph{mapping functions}. In our work, each neuron in the input layer is assigned to a service attributes $a \in A$. A single service attribute represents one of its inherent characteristics (e.g., owner's rating). Each neuron in the output layer represents a single indicator $F \in \mathcal{F}$. The goal of the training phase is to find a set of synapse weights that, when substituted in the mapping functions, maximizes the accuracy of assessing IoT services' trustworthiness for each available indicator.

The model is trained and built during the offline phase. IoT consumers use the model during the online phase to assess the trust level of IoT services at each available indicator $F \in \mathcal{F}$. The model takes the attributes $A$ of a given service $S$ and outputs the vector $\mathcal{F}_S$. Elements in the vector represent the trust levels of the given service at each trust indicator.

\begin{equation}
    \mathcal{F}_S = \mathcal{M}_{\mathcal{F}}^{\mathcal{S}}|_S
    \label{eq:f_s}
\end{equation}

\subsection{Per-indicator Expectation}
\label{section:expectation}

\begin{figure}
    \centering
    \includegraphics[width=0.40\textwidth]{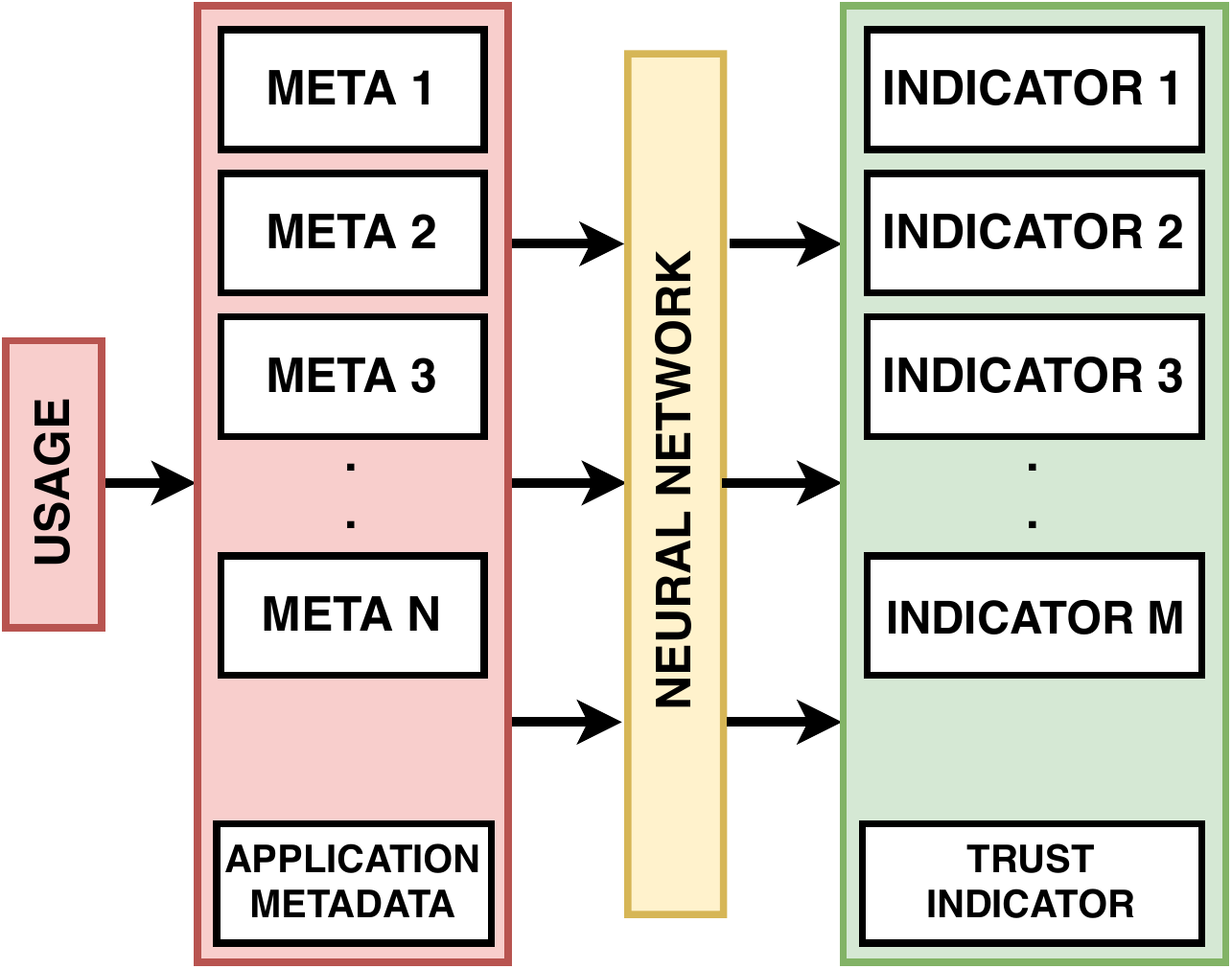}
    \caption{Usage-to-Indicator Model}
    \label{fig:usage_to_indicator_model}
\end{figure}
IoT services can be used differently by their consumers, e.g., in crowdsourced WiFi hotspot services, consumers may use the service for checking their social websites, calling, listening to music, etc. Different uses may favor different sets of trust indicators for a service to have a high trust value. Therefore, the \emph{per-indicator expectation} stage aims to predict the required trust level of each indicator given a particular usage \ref{fig:usage_to_indicator_model}. Similar to the per-indicator trust prediction stage, the per-indicator expectation's execution is split between the online and offline phase. During the offline phase, a \emph{usage-to-indicator} model $\left(\mathcal{M}_{\mathcal{F}}^U\right)$ is trained. The built model is then used during the online phase to predict the required trust level of each indicator given a consumer's usage.

We represent a usage $u$ as a set of metadata $M_u$. As mentioned before, single metadata $m \in M_u$ is a word that describes certain aspects of the usage $u$. The set of metadata cannot be used directly to train the usage-to-indicator model. Therefore, the set $M_u$ is converted into a vector. We refer to the vector as \emph{descriptor vector} $\vec{M_u}$. The length of the descriptor vector is equal to the number of all potential metadata. We obtain the $k^{th}$ element in $\vec{M_u}$ as follows:

\begin{equation}
    v(k, u) = 
     \begin{cases}
      \text{1,} &\quad\text{if } \vec{M}_k \in M_{u}\\
      \text{0,} &\text{otherwise} \\
     \end{cases}
    \label{eq:descriptor_vector}
\end{equation}
Where $\vec{M}$ is a vector containing all possible metadata and $\vec{M}_k$ is the $k^{th}$ element in the vector. As an illustration to Equation \ref{eq:descriptor_vector}, suppose a crowdsourced WiFi hotspot environment, where $\vec{M} = [streaming, finance, social, gaming]$, $u = Twitter$ , and  $M_{u} = \{social\}$. By substituting in Equation \ref{eq:descriptor_vector}, the descriptor vector $\vec{M_u}$ is $[0, 0, 1, 0]$.

Previous rating data is used to train the usage-to-indicator model. Usage descriptor vectors $\vec{M_u}$ and detected trust indicators $\mathcal{F}$ are used to train and build the model. Similar to the per-indicator trust prediction stage, Neural Networks (NN) are used for training the model. The neurons of the input layer represent the elements of the descriptor vector $\vec{M_u}$. Neurons in the output layer correspond to the trust indicators $F \in \mathcal{F}$. 

The usage-to-indicator model is used during the online phase. IoT consumers use the model to assess the required trust levels of their usage for each trust indicator. The output of the model is a vector $\mathcal{F}_u$. Each element in the vector represents the required trust level for each trust indicator:

\begin{equation}
    \mathcal{F}_u = \mathcal{M}_{\mathcal{F}}^U|_u
    \label{eq:f_u}
\end{equation}

\begin{table}[!t]
\caption{Trust expectations per usage.}\
\centering
\begin{tabular}{|l|c|c|}
\hline
Usage & Internet Speed & Security \\
\hline
Usage A & 6 & 1 \\
Usage B & 9 & 2 \\
Usage C & 9 & 3 \\
Usage D & 1 & 9 \\
\hline
\end{tabular}
\label{tab:example_expectation}
\end{table}

\subsection{Adaptive Trust Assessment}
\label{section:adaptive_trust_assessment}
Given an IoT service $S$ and consumer usage $u$, the final stage aggregates the results from previous stages to obtain a trust value \emph{adapted} to the consumer's usage. More specifically, the service's trust level at each trust indicator $\mathcal{F}_S$ and the usage's required trust level at each trust indicator $\mathcal{F}_u$ are used to evaluate the \emph{adaptive trust} $\mathcal{T}_{Adapt}(S, u)$ as follows:

\begin{equation}
    \mathcal{T}_{Adapt}(S, u) = \frac{\sum\limits_{n=1}^{n = |\mathcal{F}|} \min(\mathcal{F}_S(n), \mathcal{F}_u(n))}{\sum\limits_{f \in \mathcal{F}_u} f}
    \label{eq:t_adapt}
\end{equation}
The adaptive trust value can be between 0 and 1, where a value equals to 1 indicates that a service $S$ satisfies all trust indicators required by a specific usage $u$.

\section{Multi-Use Adaptive Trust}
\label{section:multi_usage}
We build on our earlier framework by introducing an approach that considers \emph{multiple} uses for a single IoT consumer. The approach is split into two phases: Usage Pattern Prediction, and Per-Usage Trust Indicators Aggregation. Usage pattern can vary from one IoT consumer to another. For example, one user may check their news feed, chat with their friends on WhatsApp, and then watch videos on YouTube. Another user may call their family on Skype, shop on Amazon, and pay their bills. The first phase, therefore, aims at predicting such patterns. The second phase provides an overall presentation of the predicted usage pattern. Note that each usage in the pattern has a certain trust expectation/requirement. The goal of the Per-Usage Trust Indicators Aggregation phase is to find an accurate aggregation that closely represents every per-usage trust expectation.

\begin{figure*}[!t]
\centering
\begin{minipage}{.40\textwidth}
  \centering
  \includegraphics[width=1.0\linewidth]{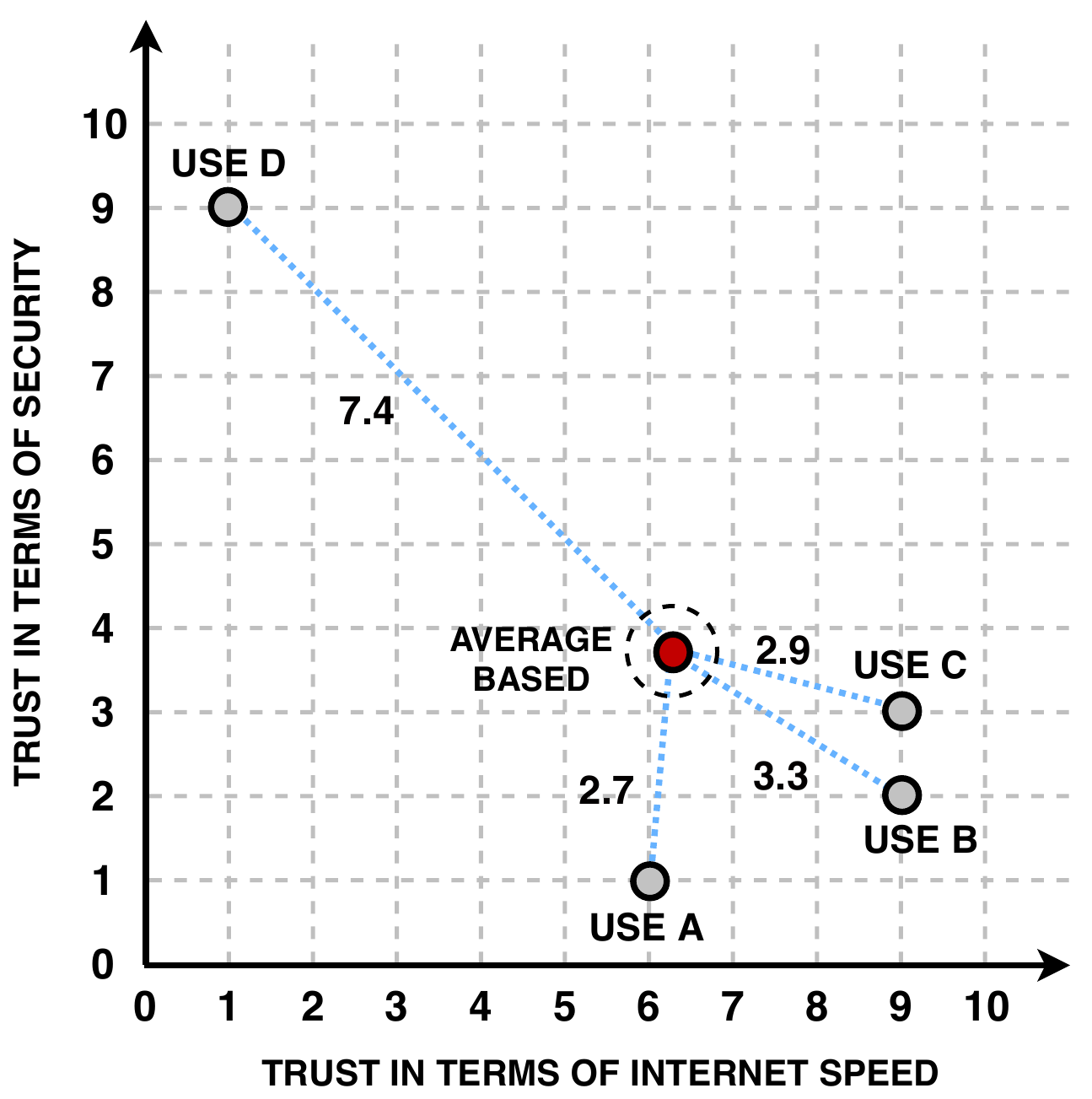}
  \caption{An example of Average-Based Aggregation}
  \label{fig:average_aggregation_example}
\end{minipage}%
\hspace{1.5cm}
\begin{minipage}{.40\textwidth}
  \centering
  \includegraphics[width=1.0\linewidth]{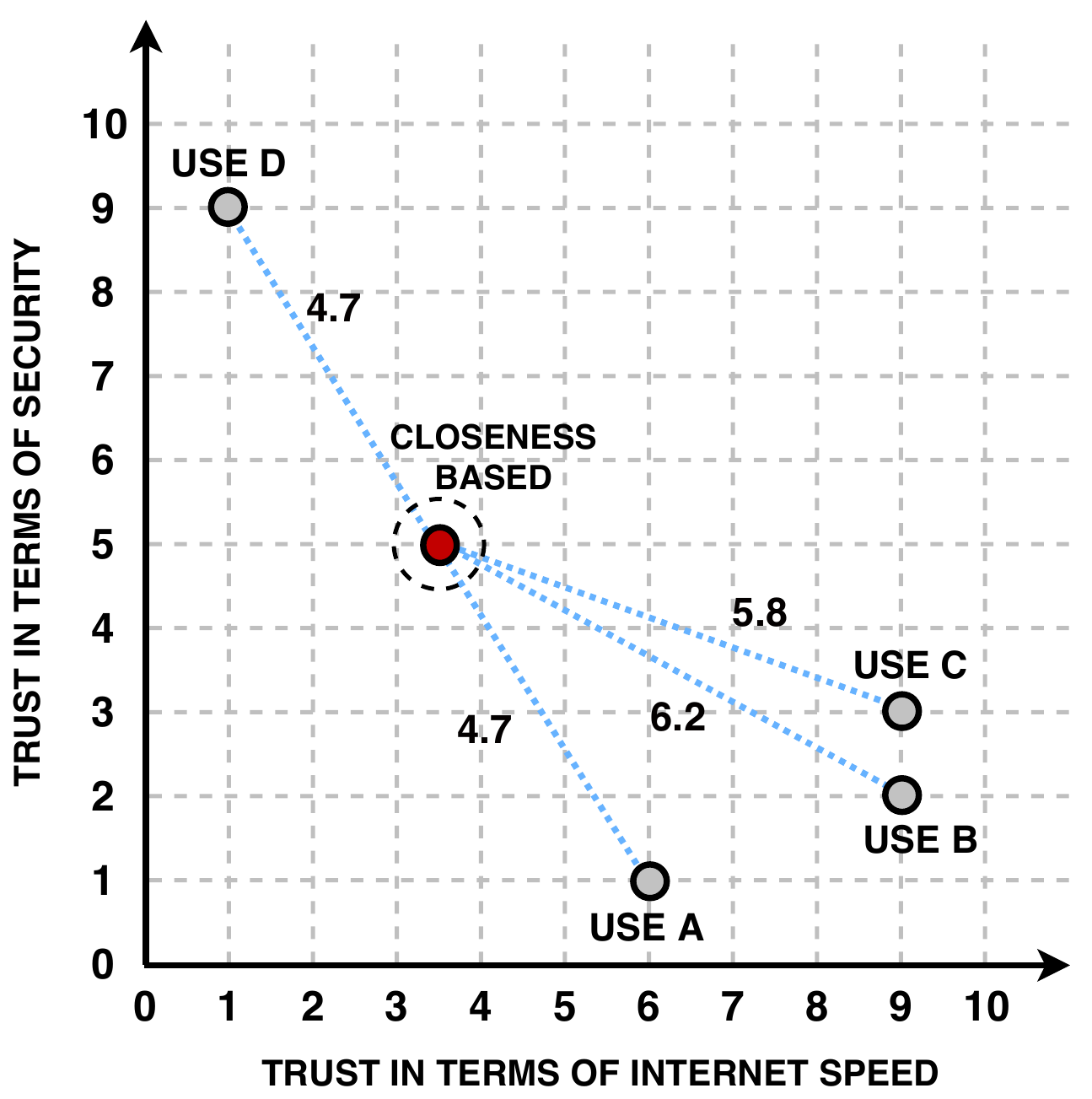}
  \caption{An example of Closeness-Based Aggregation}
  \label{fig:closeness_aggregation_example}
\end{minipage}
\end{figure*}

\subsection{Usage Pattern Prediction}
The first phase of the approach is to predict the usage pattern a consumer may have during service provision. It has been shown in the literature that IoT users exhibit unique features that can be leveraged to achieve such a goal \cite{tu2018your}  \cite{welke2016differentiating}. These features include mobility patterns, web browsing histories, and sensor's fingerprints. The work in \cite{huang2012predicting} can be used to predict consumers' usage patterns. Our choice of their work is due to the closeness of their setup to IoT environments. The prediction model is based on application uses in smartphones which resembles IoT crowdsourcing platforms. Their approach relies on contextual features to build the prediction model. Examples of these features include location, time, and previous uses. Such properties are generally manifested in IoT devices. In other words, the habits of users (mobility, current usage, time, etc.) are the primary input for the proposed prediction model. IoT devices share that property with mobile devices. A single IoT user may have the habit of using their devices in a unique way the same way smartphones users do.

The prediction process in \cite{huang2012predicting} covers two stages; training and prediction. During the training stage, relevant contextual features are extracted from raw datasets. A feature is any piece of information that can hint or influence the selection of an application by smartphone users.  The features are then compared against each other to look for any possible correlations (e.g., time and location). A model is built based on the discovered features and the similarities between them. During the prediction phase, the trained model is fed with context information and predicts the application to be used by a given user.

\subsection{Per-Usage Trust Expectation Aggregation}

An IoT consumer usage pattern can be expressed as a set of potential uses $U_c$ during a single IoT service session. We assess the expected trust values for each usage $u \in U_c$ at every trust indicator $F \in \mathcal{F}$. Note that a service's trust can be influenced by several indicators, e.g., security and Internet speed in WiFi hotspot services. Usage $A$ might expect different trust values compared to another usage $B$. The variations between usage $A$ and $B$ may happen on a per trust indicator basis. For example, one consumer might expect a WiFi service to be secure, while another expects the service to have a fast Internet speed depending on their usage. The expected trust value at each indicator by each usage is computed using our Per-Indicator Expectation discussed earlier. The result is a vector $\mathcal{F}_u$ whose elements are the expected trust values for usage $u$. The vector $\mathcal{F}_u$ is computed for each usage $u \in U_c$. The last step is to aggregate these vectors in a way that \emph{fairly} represents each usage $u \in U_c$, in other words:

\begin{equation}
    \mathcal{F}_{U_c} \mid \forall \mathcal{F}_{u \in U_c}: \mathcal{F}_{U_c} \approx \mathcal{F}_u
\end{equation}
where $\mathcal{F}_{U_c}$ is the aggregated trust expectation. We define the \emph{fairness} $P$ of an aggregation by:

\begin{equation}
    P(\mathcal{F}_{U_c}) = \frac{\min_{u \in U_c} d(\mathcal{F}_{U_c}, \mathcal{F}_{u})}
         {\max_{u \in U_c} d(\mathcal{F}_{U_c}, \mathcal{F}_{u})}
    \label{eq:fairness}
\end{equation}
where $d(.)$ is the Euclidean distance between two vectors. The fairness measure shows whether the aggregated expectation vector is close to the component expectation vectors. A value of one indicates that aggregation is equally distanced from the other expectation vectors. A lesser value suggests that some vectors are closer to the aggregation than the others. 



We propose three aggregation methods; namely, \emph{Average-Based Aggregation, Closeness-Based Aggregation}, and \emph{Usage-Significance Closeness-Based Aggregation}. We consider the following running example throughout the discussion of our aggregation methods. Assume an IoT consumer who wants to use a WiFi hotspot service. The consumer is predicted to use four applications: Use $A$, $B$, $C$, and $D$. For simplicity, the number of trust indicators is two; namely, Internet speed, and security. Assume that the trust expectations of the four uses have been evaluated using the Per-Indicator Expectation stage from our proposed adaptive trust framework. The evaluation results are shown in Table \ref{tab:example_expectation}. For instance, Use $B$ requires a WiFi service that have at least a trust value of 10 and 2 in terms of Internet speed and security, respectively. In this example, we use trust values ranging from 0 to 10 for simplicity, where 0 is the lowest trust value and 10 is the highest.

\subsubsection*{Average-Based Aggregation}
Given a set of consumer uses, the aggregation of their trust expectation can be achieved by taking the average. More formally, the Average-Based method aggregates trust expectations as follows:

\begin{equation}
    \mathcal{F}_{U_c}^{avg} = \frac{\sum\limits_{u \in U_c}\mathcal{F}_u}{|U_c|}
\end{equation}

In the previous example, the aggregated trust expectation is 6.25 for speed and 3.75 for security using Average-Based Aggregation. The individual and aggregated expectations are visualized in Fig. \ref{fig:average_aggregation_example}. The aggregated trust expectation is closer to the expectations of Usage $A$, $B$, $C$ than Usage $D$. Using such aggregation can introduce inaccurate trust assessment since Usage $D$ might not be satisfied. The fairness measure is 0.37 using equation \ref{eq:fairness}. The low value indicates that the aggregation favors certain uses over others as evident in Fig. \ref{fig:average_aggregation_example}.

\subsubsection*{Closeness-Based Aggregation}
\begin{figure*}[!t]
\centering
\begin{minipage}{.47\textwidth}
  \centering
  \vspace{2.5cm}
  \includegraphics[width=1.0\linewidth]{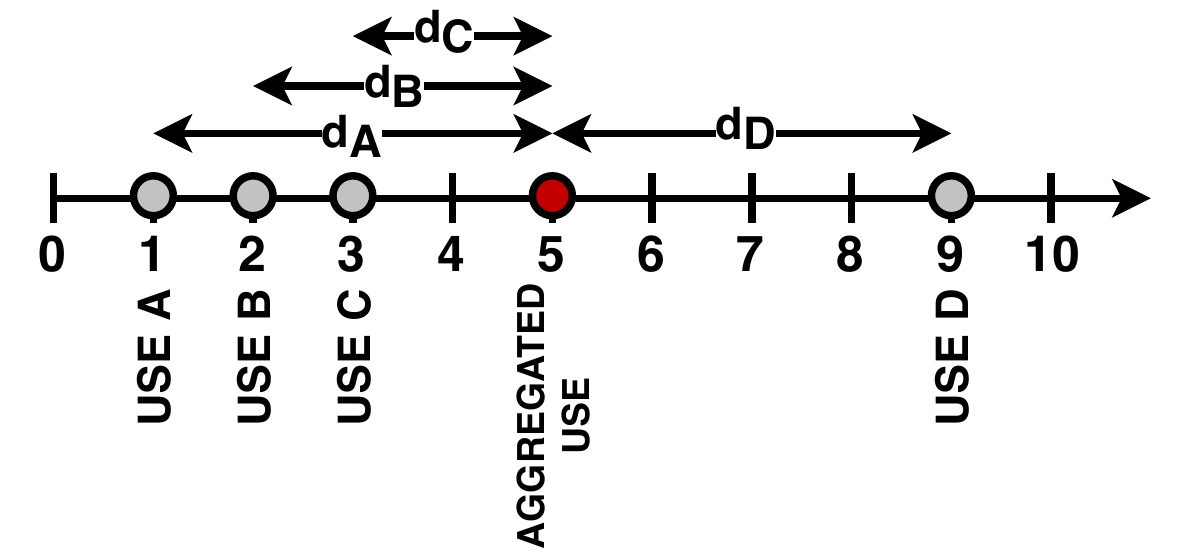}
  \vspace{0.1cm}
  \caption{An example of trust expectations in terms of service security.}
  \label{fig:one_dimensional_min_distance}
\end{minipage}%
\hspace{0.8cm}
\begin{minipage}{.39\textwidth}
  \centering
  \includegraphics[width=1.0\linewidth]{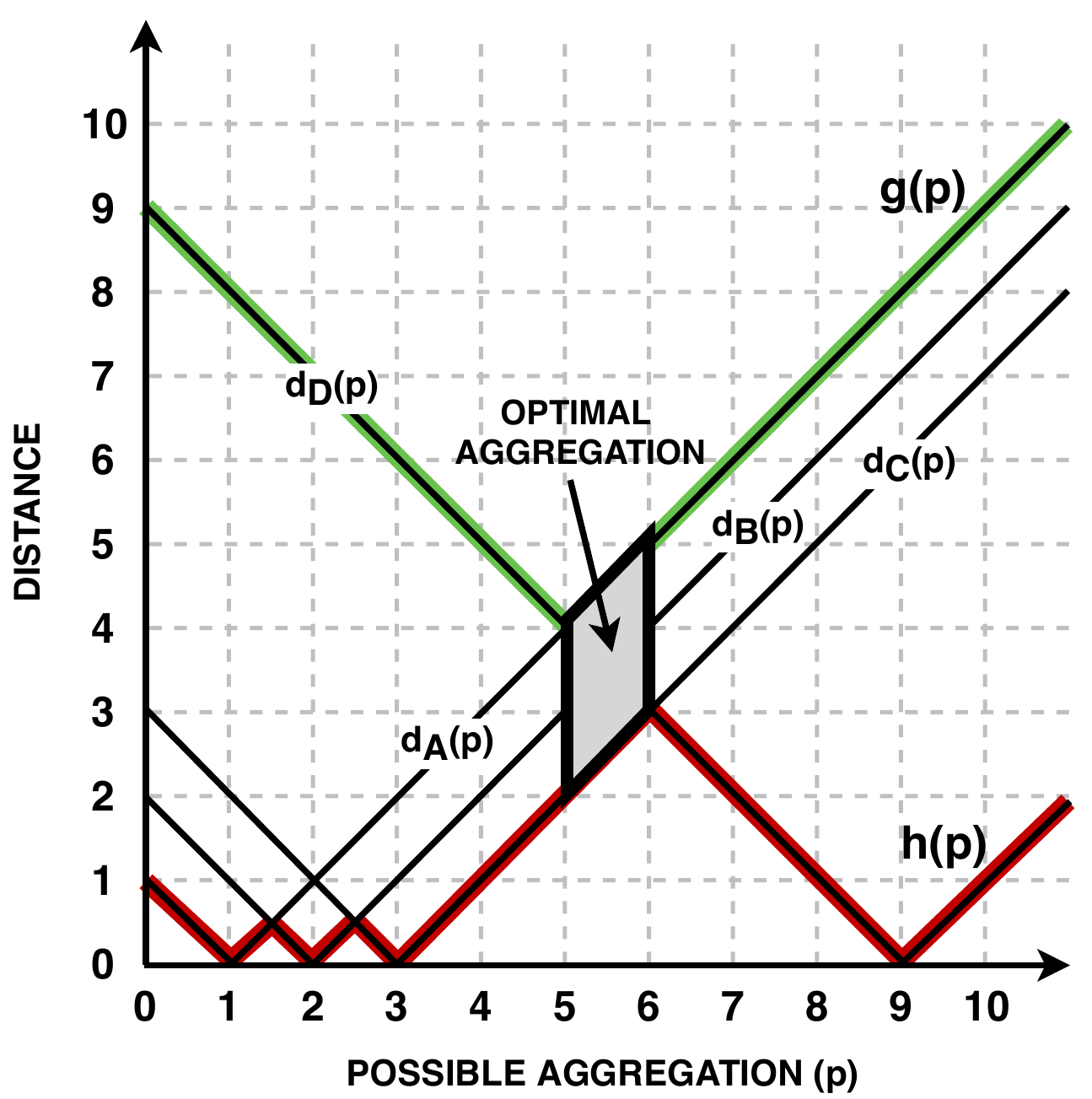}
  \caption{An example of optimal aggregation given a set of trust expectations.}
  \label{fig:optimization_example}
\end{minipage}

\end{figure*}
The intuition behind the Closeness-Based aggregation is to find an aggregation that satisfies two conditions. First, the aggregation is as close as possible to the component trust expectations. Second, the distances between the aggregation and the component expectations are as equal as they can be. Fig. \ref{fig:closeness_aggregation_example} visualizes the earlier example uses and an aggregation that satisfies the two  conditions. The aggregated trust expectation is 3.54 for Internet speed and 5.00 for security. The resulting aggregation equally satisfies all of the consumer's uses. The aggregation's fairness is 0.76, according to equation \ref{eq:fairness}; two times more than the fairness using Average-Based Aggregation.

Finding a trust expectation that satisfies the two conditions can be achieved by formulating the problem as an optimization problem. For simplicity, we focus on the trust expectation in terms of security only. Fig. \ref{fig:one_dimensional_min_distance} shows the four uses and a possible aggregation $p$. There are four distances from the possible aggregation $p$ to each usage. The aim is to minimize both the distances and the difference between them. Each distance can be rewritten as a function of the target aggregation $p$ as follows:

\begin{equation*}
    \begin{aligned}
        d_A(p) = |p - 1| \\
        d_B(p) = |p - 2| \\
        d_C(p) = |p - 3| \\
        d_D(p) = |p - 9|
    \end{aligned}
\end{equation*}
Fig. \ref{fig:optimization_example} shows plots of the four functions (black curves). The aforementioned two conditions can be satisfied if we minimize the difference between the minimum and maximum of all earlier functions. We, therefore, define two more functions as follows (visualized in Fig \ref{fig:optimization_example} in red and green):
\begin{equation*}
    \begin{aligned}
        g(p) = max(d_A(p), d_B(p), d_C(p), d_D(p)) \\
        h(p) = min(d_A(p), d_B(p), d_C(p), d_D(p))
    \end{aligned}
\end{equation*}
 In other words, the optimal aggregation $p*$ is the point $p$ where if substituted in $g$ and $h$, would give the minimum difference between the two functions, i.e.:
\begin{equation*}
    \begin{aligned}
        p* = \argmin_p \{p \mid g(p) - h(p)\}
    \end{aligned}
\end{equation*}


It is worth noting that the solution $p*$ is a range of points that satisfy the Closeness-Based Aggregation conditions as shown in Fig. \ref{fig:optimization_example} (the grey region). The following is a general optimization problem formulation for the $k^{th}$ indicator:

\begin{equation}
    \begin{aligned}
        \argmin_p \quad & \{p \mid g(p) - h(p)\} \\
        \textrm{s.t.} \quad & D = \{d(p) = |p - \mathcal{F}_u(k)|, u \in U_c\} \\
                   & g(p) = \max_{d \in D} d(p) \\
                   & h(p) = \min_{d \in D} d(p)
    \end{aligned}
    \label{eq:optimization}
\end{equation}

The trust expectations are trust values ranging from zero to one. As a result, the aggregation has the same range. Additionally, a practical trust value can have two decimal points for an accurate assessment. Increasing the precision beyond two decimal points does not increase the accuracy significantly. As a result, the \emph{search space}, where the optimal solution lies, has a range of $[0, 1]$. There are 101 intermediate values in the search space using our two-decimal precision (i.e., 0.00, 0.01, 0.02 ..., 1.00). We opt for the exhaustive search to solve the optimization problem because of the limited search space.

\subsubsection*{Usage-Significance Closeness-Based Aggregation}
In some scenarios, a particular usage might be more important than other uses. Trust expectation aggregation should take into consideration such cases. For example, a consumer might consume a WiFi service to use two applications $A$ and $B$ on their smartphones. Application $A$ might be used over a longer period than the application $B$. In such a scenario, the \emph{duration} can be one of the attributes that affects the usage's significant. Obtaining the duration an application is fairly straightforward since modern operating systems offer built-in tools for that. We augment the optimization problem listed in equation \ref{eq:optimization} with the \emph{significance} of the usage as follows:

\begin{equation}
    \begin{aligned}
        \argmin_p \quad & \{p \mid g(p) - h(p)\} \\
        \textrm{s.t.} \quad & D = \{d(p) = \mathcal{S}_u|p - \mathcal{F}_u(k)|, u \in U_c\} \\
                   & g(p) = \max_{d \in D} d(p) \\
                   & h(p) = \min_{d \in D} d(p)
    \end{aligned}
    \label{eq:optimization_significance}
\end{equation}
where $\mathcal{S}_u$ is the significance of the usage $u$. The significance is a value that ranges from zero to one, one being highly significant. Several usage attributes can contribute to its significance. In this paper, we focus on the \emph{duration} attribute. In other words, the longer the usage period, the more significant it is. We formulate the significance of the usage $u$ as follows:

\begin{equation}
    \mathcal{S}_u = \frac{dr_u}{\sum\limits_{u \in U_c} dr_u}
\end{equation}
where $dr_u$ is the average duration of the usage $u$. It is worth noting that the significance of a usage is relative to the other uses in $U_c$. The summation of all significance, therefore, should be equal to one.

\begin{figure*}[!t]
\centering
\begin{minipage}{.48\textwidth}
    \centering
    \includegraphics[width=1.0\textwidth]{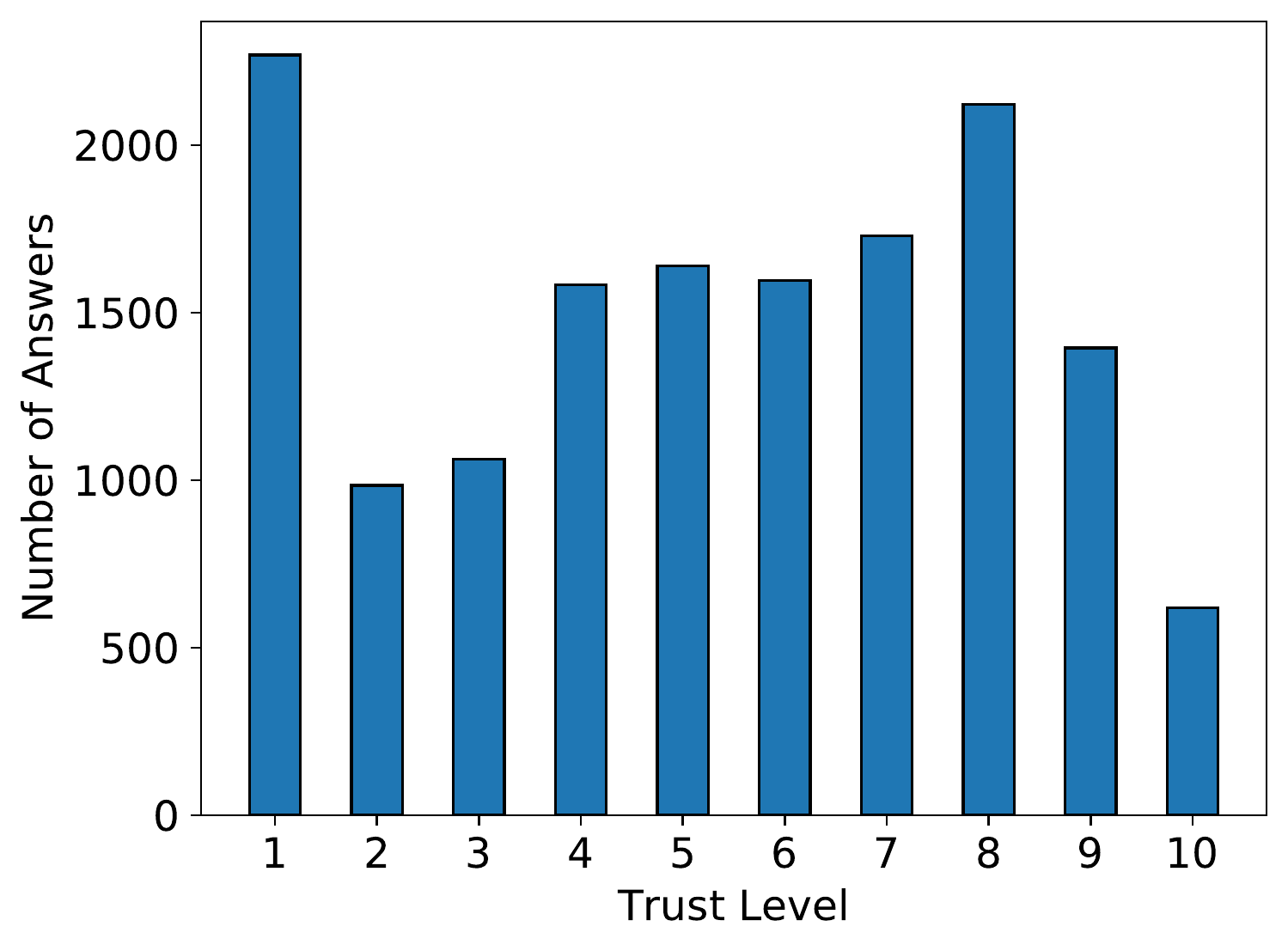}
    \caption{The distribution of mTurk workers' answers}
    \label{fig:answers_distribution}
\end{minipage}%
\hspace{0.5cm}
\begin{minipage}{.48\textwidth}
    \centering
    \includegraphics[width=1.0\textwidth]{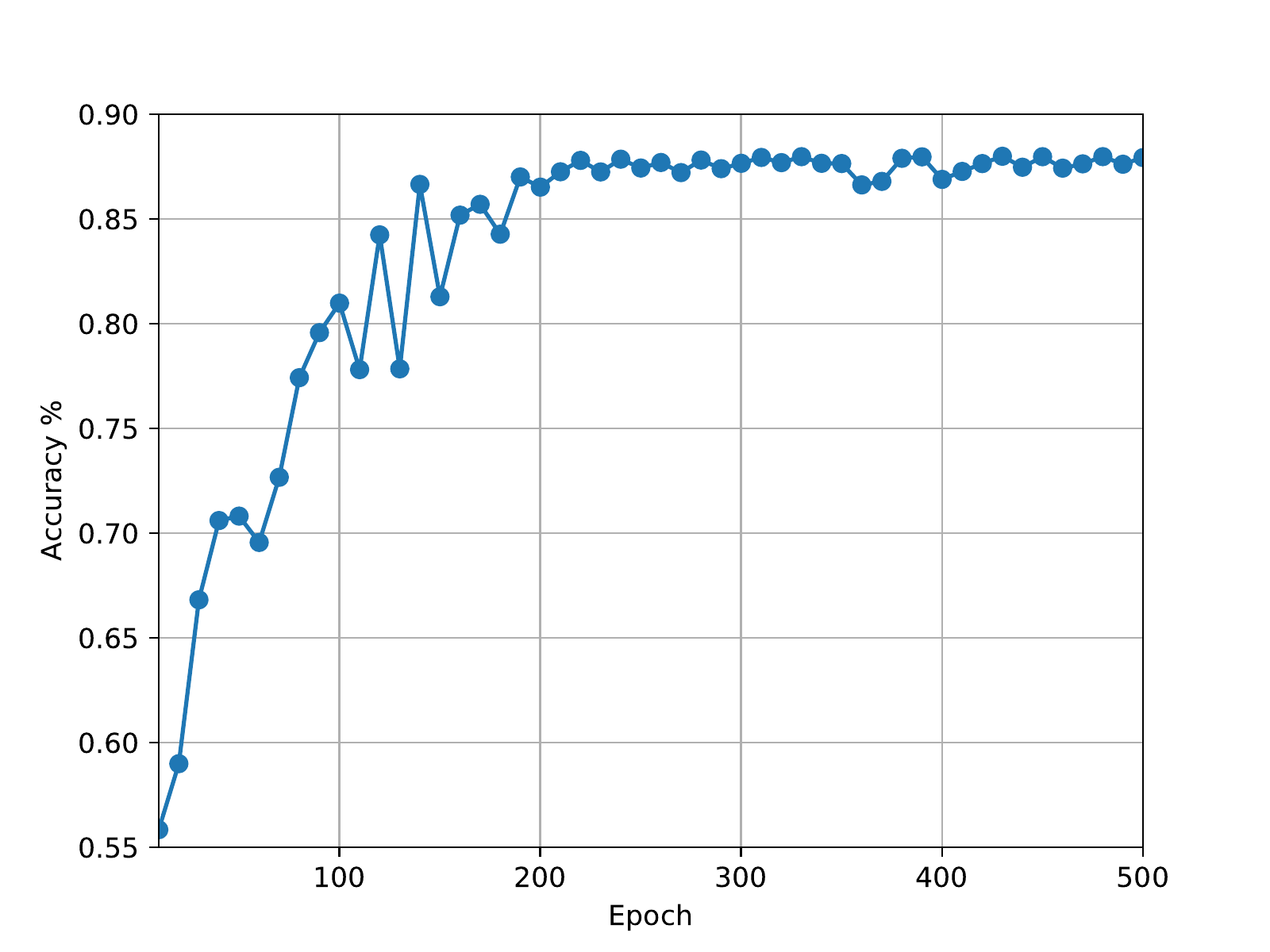}
    \caption{Framework accuracy (Single Usage).}
    \label{fig:accuracy}
\end{minipage}
\end{figure*}

\section{Evaluation}
\label{sec:evaluation}
We conduct two sets of experiments to evaluate the performance of the proposed approach. The first set examines the accuracy of the proposed framework in assessing the trustworthiness of IoT services for several uses. The second set of experiments investigates the framework's training time. We run our experiments on a 3.60GHz Intel(R) Core(TM) i7-7700 and 8 GB of RAM.

\subsection{Dataset Description}
We use the crowdsourcing platform Amazon Mechanical Turk\footnote{https://www.mturk.com/} (MTruk) to collect the dataset for our experiments\footnote{This dataset was collected due to the absence of any publicly available IoT crowdsourcing environment dataset.}. Several questionnaires are presented to Mechanical Turk workers. Each questionnaire starts by describing the IoT crowdsourcing environment to workers. We adopt a crowdsourced WiFi hotspot environment where, given a specific area, users can share/use WiFi hotspots to/from other IoT devices. MTurk workers are asked to consider themselves as service consumers. Additionally, MTurk workers are asked to assume that a WiFi hotspot service is available. A WiFi service is portrayed as \emph{a list of attributes}. The attributes used in the questionnaires include \emph{owner reputation}, \emph{device brand}, \emph{device model}, \emph{device operating system}, and \emph{carrier reputation}. Workers are asked to assume that they will use the provided service for a specific \emph{usage}. Uses include (but not limited to): watching YouTube videos, shopping on Amazon, voice-calling using Skype, and searching on Google. Workers are asked to use their best judgment to assess the trustworthiness of the given service \emph{based on how the service is used}. Additionally, each worker is presented with a list of three uses. The workers are asked to assign an overall trust value for the service given the three uses. The three uses are used in our experiments as the \emph{usage pattern} for a given service consumer. Workers submit their assessment by giving a value between 1 and 10, 1 indicates an untrustworthy service, whereas 10 represents a highly trusted service. A total of 15000 questionnaires are created and published on MTurk.  Each questionnaire presents a new service consumption instance by varying service attributes and usage. The distribution of the workers answers is shown in Fig. \ref{fig:answers_distribution}.

\begin{figure*}[!t]
\centering
\begin{minipage}{.48\textwidth}
    \centering
    \includegraphics[width=1.0\textwidth]{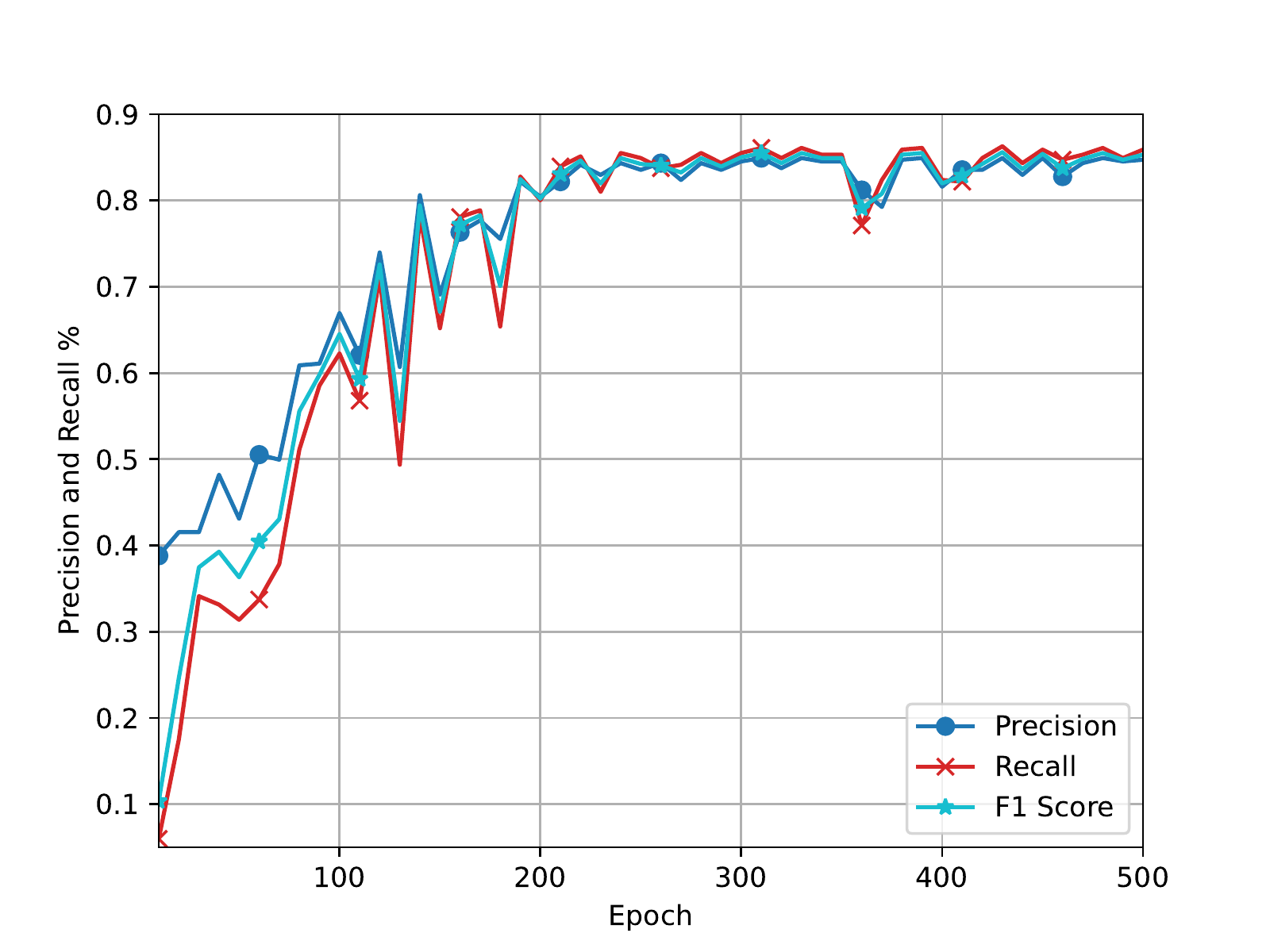}
    \caption{Framework precision, recall, and F1 scores (Single Usage).}
    \label{fig:precision_recall}
\end{minipage}%
\hspace{0.5cm}
\begin{minipage}{.48\textwidth}
    \centering
    \includegraphics[width=1.0\textwidth]{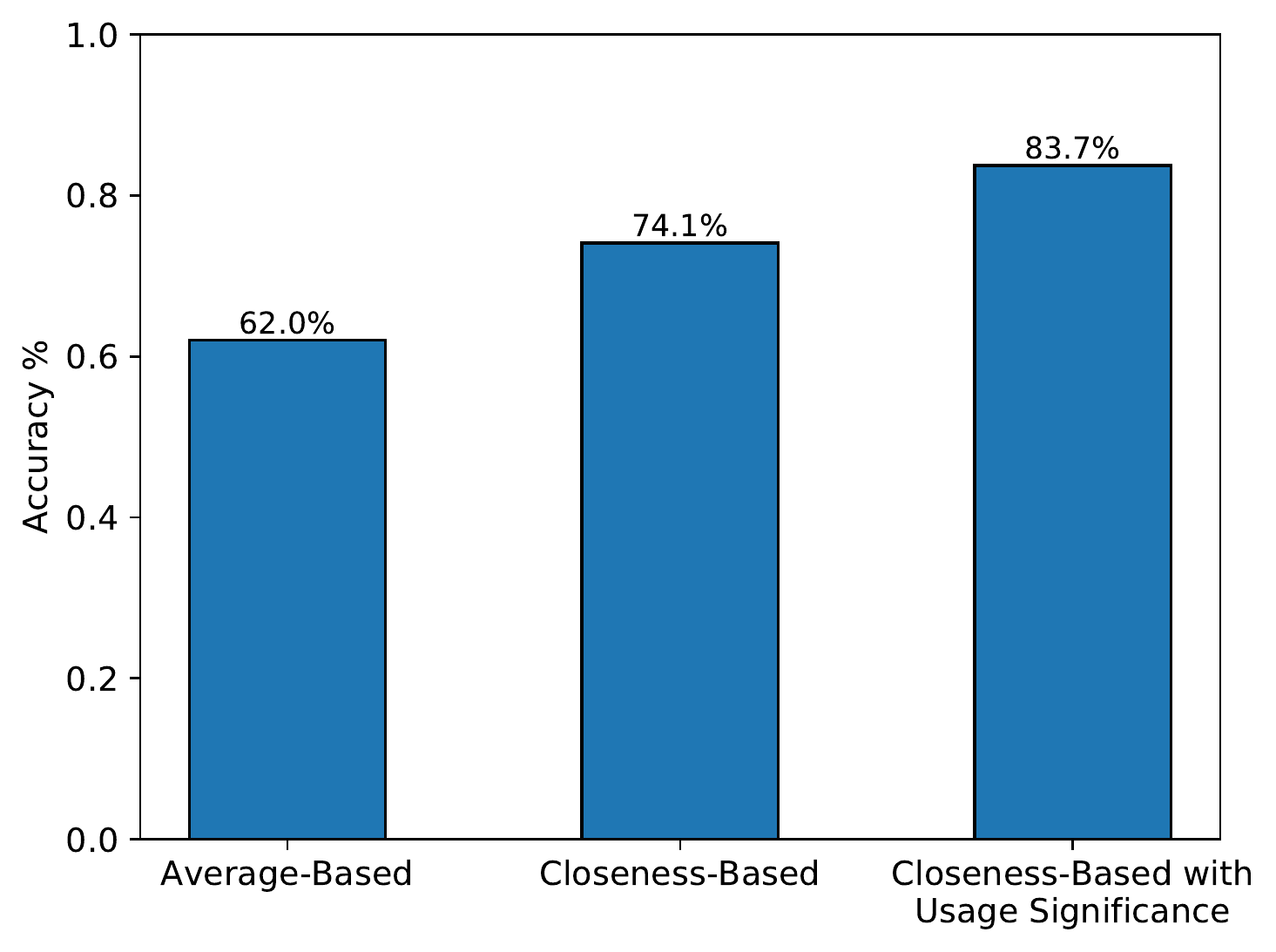}
    \caption{Accuracy for the proposed framework (Multi-Usage).}
    \label{fig:experiment_extension_accuracy}
\end{minipage}
\end{figure*}

\begin{figure*}[!t]
\centering
\begin{minipage}{.48\textwidth}
    \centering
    \includegraphics[width=1.0\textwidth]{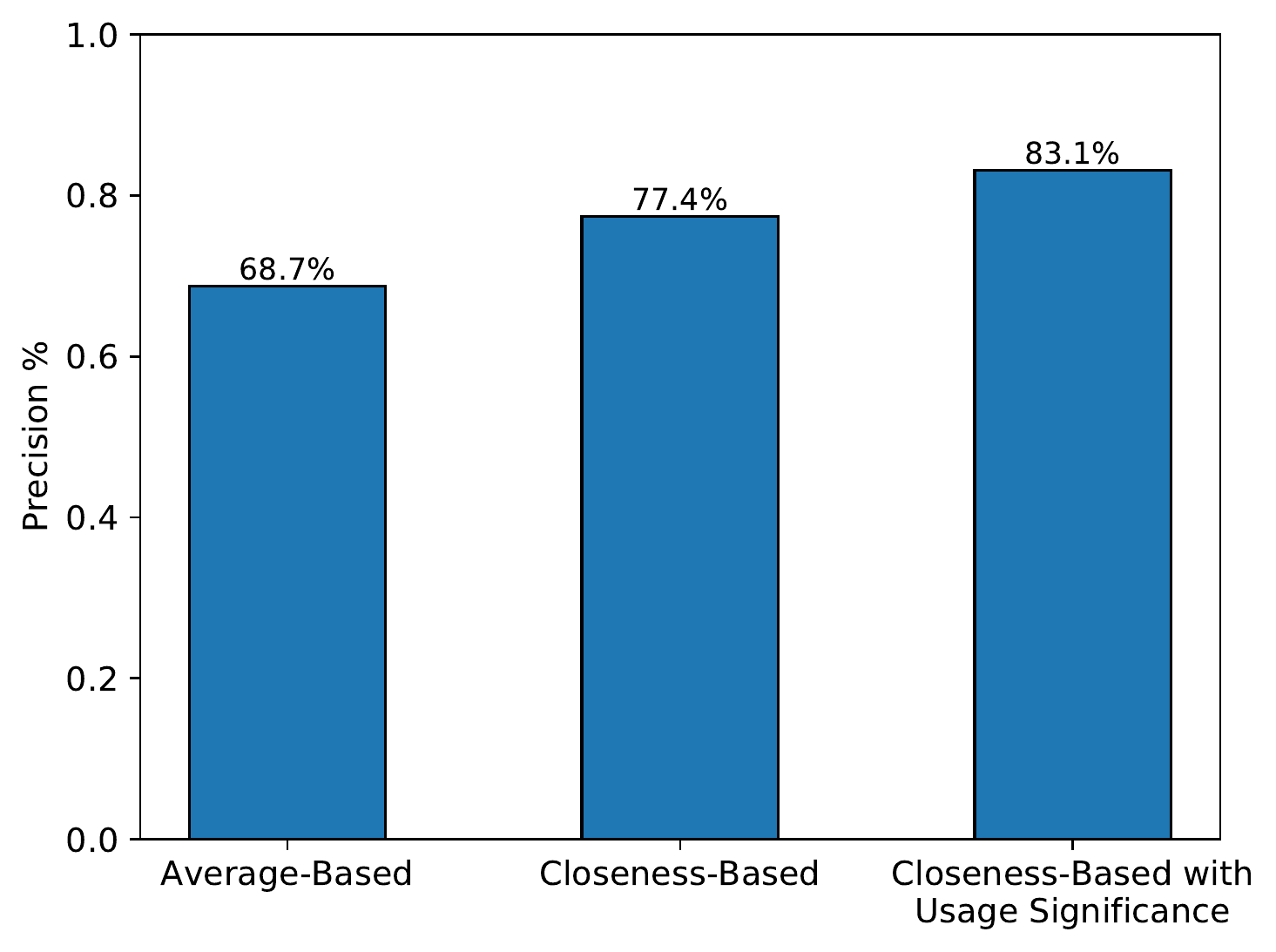}
    \caption{Precision scores for the proposed framework (Multi-Usage).}
    \label{fig:experiment_extension_precision}
\end{minipage}%
\hspace{0.5cm}
\begin{minipage}{.48\textwidth}
    \centering
    \includegraphics[width=1.0\textwidth]{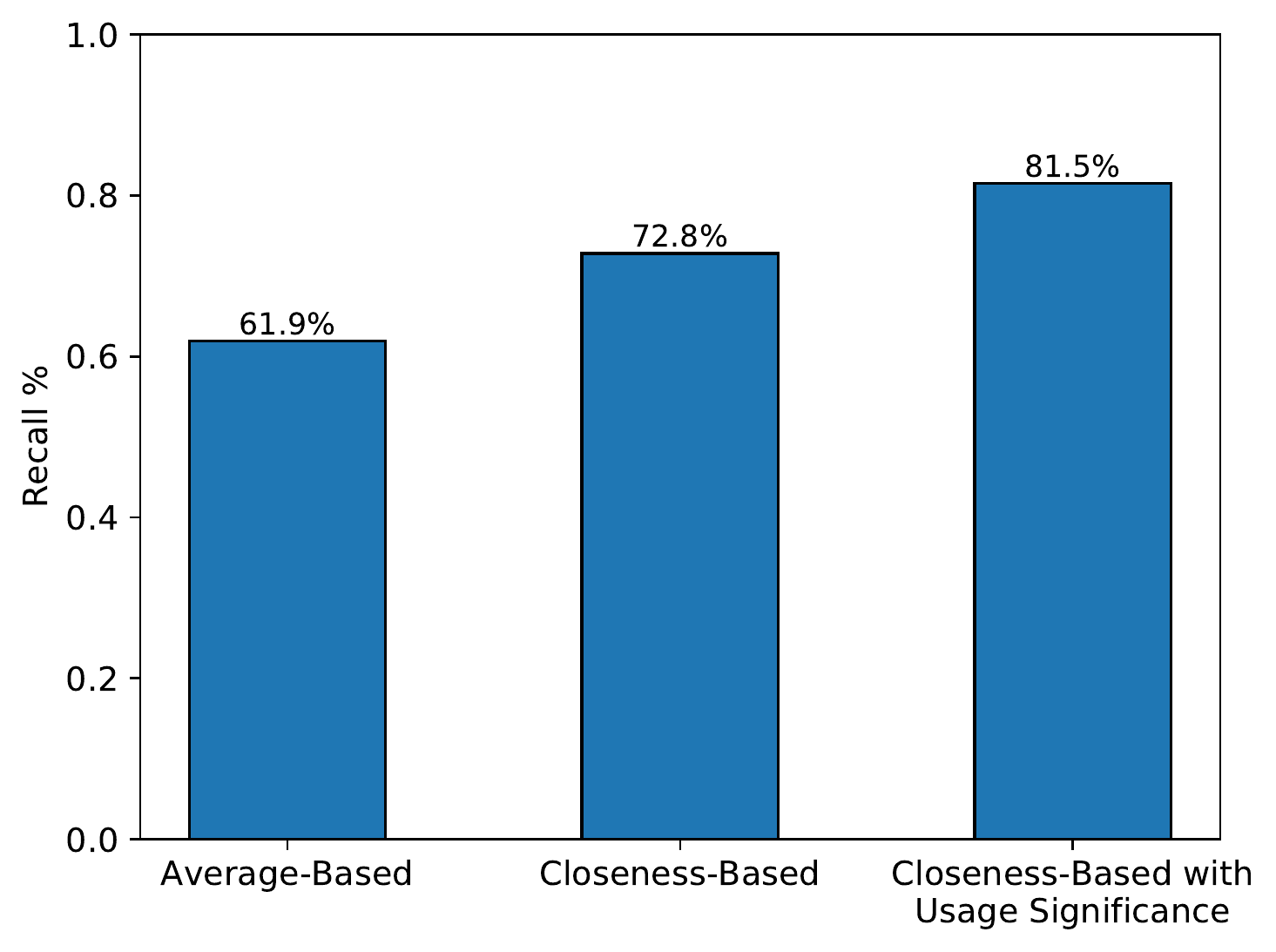}
    \caption{Recall scores for the proposed framework (Multi-Usage).}
    \label{fig:experiment_extension_recall}
\end{minipage}
\end{figure*}

\subsection{Experimental Results}

The collected data is split into two sets: (1) 80\% is used for training the framework's models, and (2) the remaining 20\% is used to compute their accuracy. 

\emph{Precision}, \emph{recall}, \emph{F1 score}, and \emph{accuracy} \cite{Olson:2008:ADM:1795943} are computed to evaluate the performance of the framework. Given a set of  samples and a specific trust level $l$ (e.g., highly trusted) the \emph{precision} for $l$ is computed as the ratio between the number of correctly detected samples as $l$ to the total number of samples detected as $l$ as follows:

\begin{equation}
    \label{eq:precision}
    Precision_l =\frac{|correct_l|}{|detected_l|}
\end{equation}

\emph{Recall} for $l$ is the ratio between the number of correctly detected samples as $l$ to the actual number of samples under $l$ in the dataset:

\begin{equation}
    \label{eq:recall}
    Recall_l =\frac{|correct_l|}{|actual_l|}
\end{equation}

\emph{F1 Score} for $l$ is the harmonic mean of the precision and recall for $l$, and can be obtained as follows:

\begin{equation}
    \label{eq:f1_score}
    F1_l = 2 \times \frac{Precision_l * Recall_l}{Precision_l + Recall_l}
\end{equation}

\emph{Accuracy} is the ratio between the number of correctly detected samples as $l$ and correctly detected samples as not $l$ to the total number of samples:

\begin{equation}
    \label{eq:accuracy}
    Accuracy_l =\frac{|correct_l|+|correct\_not_l|}{|samples|}
\end{equation}

We first evaluate the accuracy of our framework in assessing a service's trust based on a \emph{single usage}. In our experiments, we trained our models to measure the adaptive trust level of each service given a certain usage and map it to one of the 10 trust levels. Figures \ref{fig:accuracy} and \ref{fig:precision_recall} show the average precision, recall, F1 score, and accuracy in predicting the trustworthiness of IoT services. The framework achieves a high overall accuracy score of about \textasciitilde{87.9\%}. Precision, recall, F1 scores are also high, \textasciitilde{84.7\%}, \textasciitilde{85.8\%}, \textasciitilde{84.7\%}, respectively.



We evaluate the framework when consumers use services for \emph{multiple purposes} (multi-usage case). Fig. \ref{fig:experiment_extension_accuracy} shows the framework's accuracy results. The framework achieves different scores based on the used aggregation method. The highest scores are achieved when the framework uses the Closeness-Based Aggregation with Usage Significance of about 83.7\%. Using the Closeness-Based Aggregation and Average-Based Aggregation achieved 74.1\% and 62.0\%, respectively. Similarly, the precision (Fig. \ref{fig:experiment_extension_precision}) of the framework is the highest when using the Closeness-Based Aggregation with Usage Significance and the lowest when the Average-Based Aggregation is used. The framework's precision is 68.7\%, 77.4\%, and 83.1\% using Average-Based, Closeness-Based, and Closeness-Based with Usage Significance. Finally, the recall and F1 scores (Fig. \ref{fig:experiment_extension_recall}) for the framework follow a similar trend. Using the Closeness-Based with Usage Significance Aggregation the framework achieved a recall and F1 scores of about 81.5\% and 82.3\%, whereas, with Closeness-Based and Average-Based Aggregations, the recall scores are 72.8\% and 61.9\%, respectively, and the F1 scores are 65.1\% and 75.0\%, respectively.

The framework exhibited low scores whenever the Average-Based Aggregation is used. The low scores can be attributed to the fact that the aggregated trust expectation may not satisfy all consumers' uses. In other words, some uses are close to the aggregation, and some are not. As a result, the overall assessed trust value might not represent the true trustworthiness of IoT services. In contrast, the Closeness-Based Aggregation selects an aggregated trust expectation that highly satisfies all uses equally. Consequently, the framework is able to detect the services' trust values with relatively high accuracy.

\section{Related Work}
\label{relatedWork}
The evaluation of trust in IoT crowdsourcing platforms is relatively new. The majority of the work done in the literature can be grouped into two main categories: \emph{Trust Assessment based on Social Networks} and \emph{Trust Assessment based on Previous Experiences}.

\begin{figure}
    \centering
    \includegraphics[width=\linewidth]{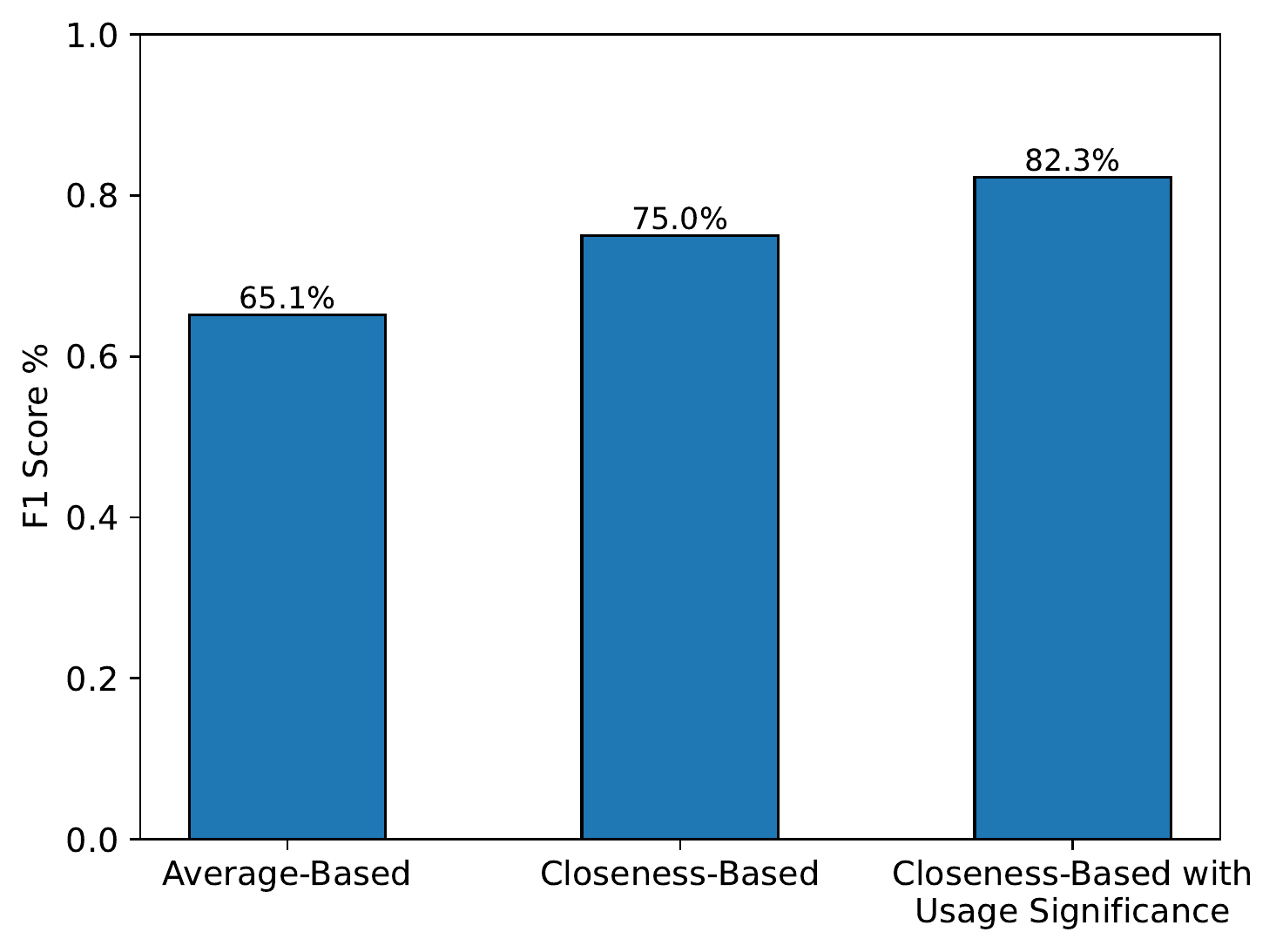}
    \caption{F1 scores for the proposed framework (Multi-Usage).}
    \label{fig:experiment_extension_f1_score}
\end{figure}


\subsection{Trust Assessment based on Previous Experiences}

A trust model is proposed for evaluating the trustworthiness and reputation of nodes in Wireless Sensor Networks (WSNs) \cite{chen2011trm}. The node's reputation is computed based on its performance characteristics: packet delivery, forwarding ratio, and energy consumption. The reputation is later used to evaluate the trustworthiness of the node. The trust can be of two types: direct and indirect. Computing the direct trust involves the trustor experience with the trustee node (using the trustee's reputation). Indirect trust is used whenever direct trust cannot be calculated due to the lack of data. The direct trust of the neighbor nodes is used to evaluate the indirect trust. The proposed model uses WSN-specific characteristics, e.g., packet delivery. This prevents the model from being used in IoT applications other than WSNs (i.e., lack of generality).

Social IoT network is used to measure the trust between two IoT devices in \cite{nitti2012subjective}. A social IoT network is a type of social networks where nodes are the IoT devices. Relationships between IoT devices indicate one or more of the following relations: similar owner, co-location, co-work, social relation, or brand. Each node computes the trust of its friends. The trust consists of a direct and an indirect trust. The direct trust is computed based on previous experiences and the type of the relation between the two IoT devices. The indirect trust is measured using the friends of an IoT device and their credibility. A trust model is also proposed in \cite{wang2016toward} that uses social IoT to manage the interactions between IoT service consumers and providers. The social IoT network is also used to search for candidate service providers. Reputation-Based trustworthiness is used to evaluate the service providers. The reputation of the provider is evaluated by assessing the results of the tasks requested by service consumers. 

A trust management protocol is proposed in \cite{bao2012dynamic, bao2012trust} to assess the trustworthiness of IoT devices. Three trust characteristics are considered: honesty, cooperativeness, and community-interest. Honesty is measured based on the direct observation of an IoT device (high recommendation discrepancy, delays, etc.). Cooperativeness and community-interest are computed using data from social networks. Common friends between two IoT owners indicate high cooperativeness between their IoT devices. Community-interest depends on the number of common communities between two IoT owners. Another framework is proposed in \cite{qureshi2020trust} to evaluate the trustworthiness among IoT nodes. The framework utilizes a Cumulative Trust Evaluation based Efficient Technique (CTBET). CTBET employs a cumulative trust concept, where the total trust is computed based on the direct and indirect trust values.

A framework is proposed for crowdsourcing services to IoT devices based on their mobility and trustworthiness \cite{kantarci2014mobility}. A central authority exists to manage interactions between the service consumer and provider. The trustworthiness of a service provider is computed based on their reputation. Basically, when the central authority receives a task request from the consumer, the task is submitted to multiple service providers. The server then computes the anomalies among the results from the service providers. Service providers with deviated results are marked and their reputation is decreased. The inclusion of a centralized authority may not be practical in IoT environments. Many devices are being deployed continuously. A centralized server can get easily overloaded by the sheer amount of IoT service requests. In addition, the framework does not address how the trustworthiness is evaluated in situations where the service provider is used for the first time, i.e., reputation bootstrapping. Another framework is proposed in \cite{wang2021mobile} that leverages the \emph{trust transitivity} to evaluate trust. An approach is designed to calculate the degrees of trust for multiple trust chains. The work also proposed an improved Dijkstra algorithm to collect trust information from nodes by mobile edge nodes. 

An approach is proposed in \cite{zhou2021truthtrust} to evaluate the trustworthiness of workers in crowdsourcing platforms such as Amazon Mechanical Turk. The approach specifically targets collusion attacks in crowdsourcing platforms. Therefore, the proposed framework leverages a trust model that is based on the CRH framework to ensure the authenticity of the evaluated trust value. Another approach is proposed in \cite{wang2020blockchain} to evaluate trust values in vehicular crowdsourcing networks (VCNs). The paper highlights that existing VCNs rely on third-party centralized trust management solutions, which might not be practical in such environments. Therefore, the paper proposes a blockchain-based decentralized framework for managing trust in VCNs. The framework consists of a trust evaluation model that assesses the trustworthiness using raters. In addition, the framework stores trust information in the blockchain for future assessments. The work in \cite{feng2020anonymous} highlighted the trade-off between preserving anonymity and trust assessment. To that end, the paper proposes a new approach that leverages Intel Software Guard Extension (SGX), which offers the ability to execute an algorithm without revealing the content of the processed data. The proposed approach assesses the trust of actors based on subjective feedback and objective behaviors.

Aforementioned approaches use historical data (i.e., previous experiences) to asses the trust. On the other hand, one key characteristic of crowdsourced IoT service environments is their high dynamism in terms of IoT devices deployment. Every day a large number of IoT services are being added. Newly added services do not have previous records. If those previous experiences are missing, the evaluated trust cannot be accurate. Therefore, these approaches cannot be utilized to accurately measure IoT services' trust.

\subsection{Measuring Trust Using Social Networks}

A social compute cloud framework is proposed in \cite{Caton2014} where users in a social network can share and consume services from other users. Some of the main issues in service provisioning are highlighted as follows: trustworthiness, reliability, and availability. The framework tries to overcome these issues by giving users the control over who can use their services and which services they can use (i.e., setting their preferences). The framework leverages the social structure of the social network. The relation types between users (e.g., family, close friends, colleagues, etc) are also utilized to determine the level of trust between them. 

A framework, namely Social WiFi, is proposed in \cite{Cao2015} that aims at eliminating the privacy risks accompanied with public WiFi hotspots. An assumption has been made that friends in social networks have a mutual trust between them. Social WiFi utilizes this trust to match hotspot users to trusted hotspot providers. The key part of the work is the integration of social WiFi into the implementation level of the WiFi standard. Specifically, WiFi network discovery and authentication is carried out while considering the social status of the two ends. The proposed framework lacks generality as it can only be used for WiFi hotspot services.

An approach for evaluating the trust between users in social networks is presented in \cite{adali2010measuring}. Behavioral interactions are used to indicate the level of trust (i.e., conversations between users and message propagation). A conversation between two users can indicate a higher level of trust if: (1) it happens many times, (2) it lasts for a long duration, and (3) there is a balanced contribution of messages from both users. The message propagation indicates the willingness of a user $B$ to forward a message received by another user $A$. A large number of forwarded messages reflect a higher trust value for the sender.

The above work focuses on social networks' relationships which is not sufficient to evaluate the trust between the IoT service provider and consumer. For example, two friends on a social network do not necessitate a mutual trust between them \cite{sherchan2013survey}. As a result, we investigate to consider other trust-related factors such as the reliability of the IoT service, and the reputation of the IoT service provider and consumer to assess the trust of crowdsourced IoT services.

\section{Conclusion}
\label{sec:conclusion}
A framework was proposed that measures the trust value of an IoT service that adapts to consumer uses. We proposed the notion of \emph{trust indicators}, a set of aspects that affect the overall trustworthiness of a service. A novel algorithm was devised to detect the trust indicators given a specific type of IoT service. A model was presented that detects the trustworthiness of a given service at each trust indicator. Another model was proposed that predicts the expected trustworthiness at each indicator given a particular usage. The models' results are then aggregated to provide a trust value tailored for a specific usage given a particular service. Additionally, we leverage the \emph{usage pattern} of IoT consumers to assess the trust of a given IoT service. Three methods were proposed to provide a representation of multiple uses by a specific consumer during a single consumption session. The framework achieved a relatively high accuracy score in our experiments.

\section{Acknowledgements}
This research was partly made possible by ARC Discovery grant DP220101823 and ARC LIEF grant LE220100078. The statements made herein are solely the responsibility of the authors.

\bibliographystyle{IEEEtran}
\bibliography{ref}

\begin{IEEEbiography}[{\includegraphics[width=1in,height=1.25in,clip,keepaspectratio]{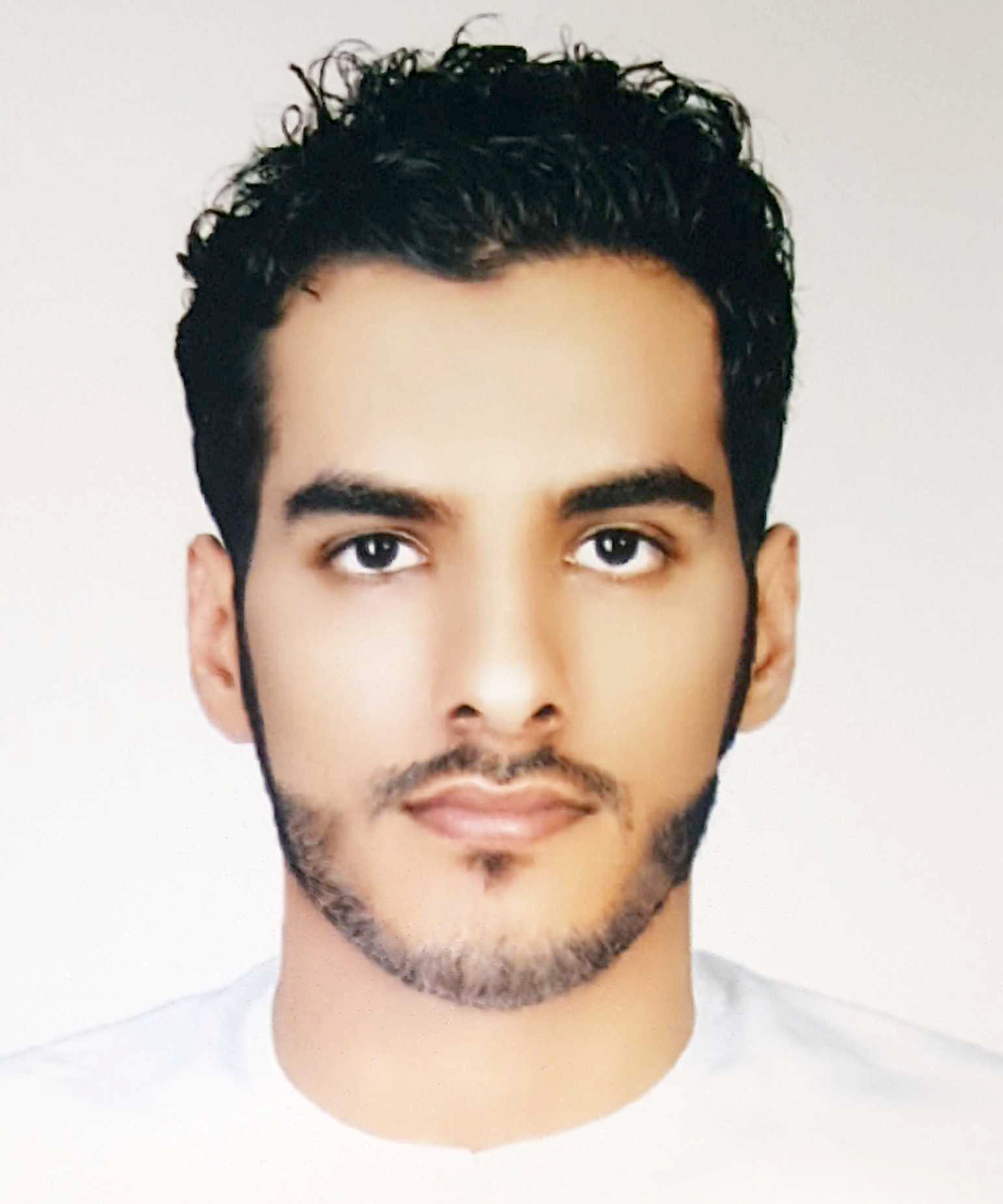}}]{Mohammed Bahutair}
is a PhD student in the School of Computer Science at  the University of Sydney, Australia. He received his bachelor degree in Computer Engineering from Ittihad University, UAE 2012 and his Masters degree in Computer Engineering from University of Sharjah, UAE 2015. His research interests are Machine Learning , Trust,  IoT and Big Data Mining.
\end{IEEEbiography}

\begin{IEEEbiography}[{\includegraphics[width=1in,height=1.25in,clip,keepaspectratio]{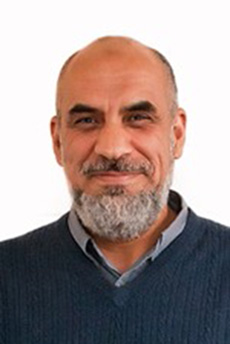}}]{Athman Bouguettaya}
is a professor at the University of Sydney,
Sydney, Australia.He received his PhD in Computer Science from the University of Colorado at Boulder (USA) in 1992. He is or has been on the editorial boards of several journals including, the IEEE Transactions on Services Computing, ACM Transactions on Internet Technology, the International Journal on Next Generation Computing, VLDB Journal, Distributed and Parallel Databases Journal, and the International Journal of Cooperative Information Systems. 
He has published more than 200 books, book chapters, and articles in journals and conferences in the area of databases and service computing. He is a Fellow of the IEEE and a Distinguished Scientist of the ACM.
\end{IEEEbiography}

\begin{IEEEbiography}[{\includegraphics[width=1in,height=1.25in,clip,keepaspectratio]{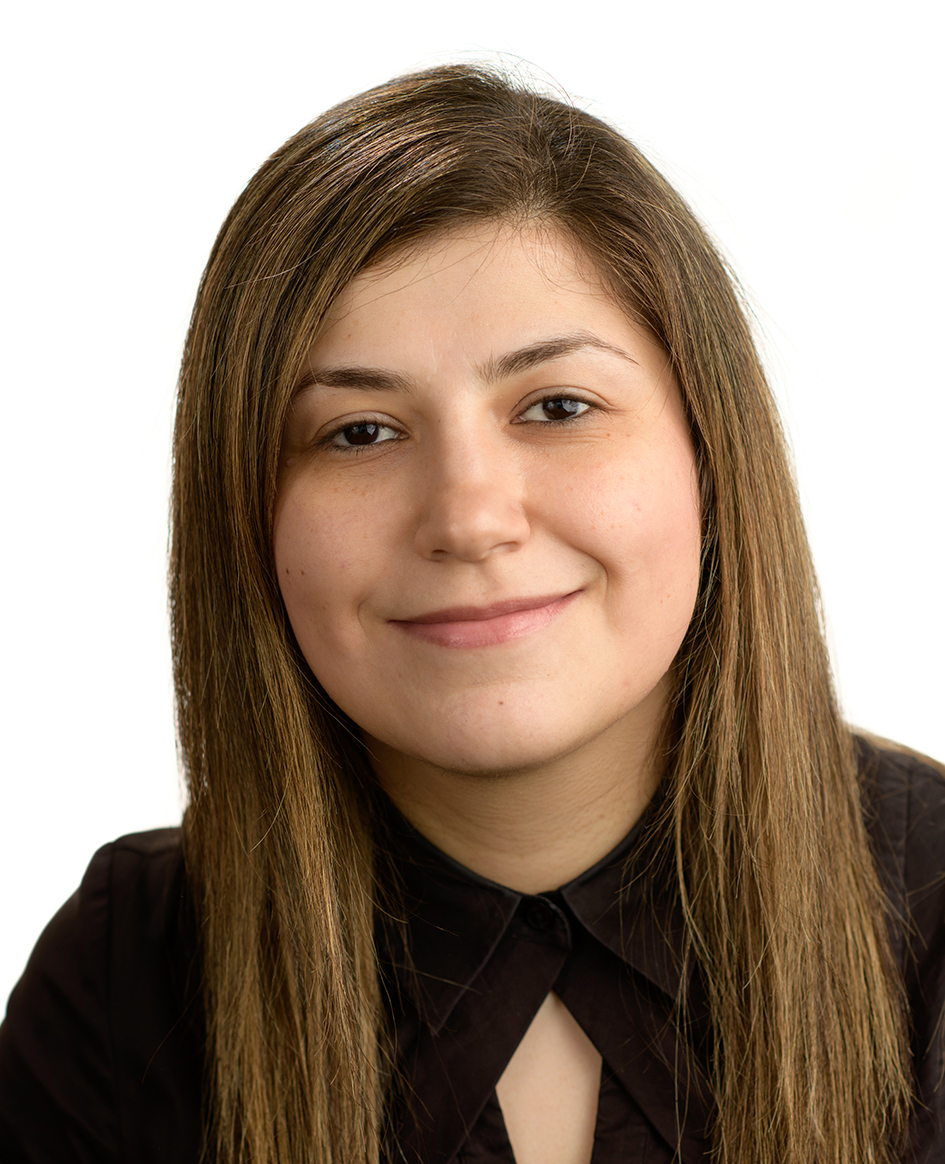}}]{Azadeh Ghari Neiat}
is a lecturer in the Information Technology at Deakin University. She was awarded a PhD in computer science at RMIT University, Australia in 2017. Her current research interests include Internet of Things (IoT), Spatio-Temporal Data Analysis, Crowdsourcing/Crowdsensing, Big Data Mining, and Machine Learning with applications in the smart city, smart home, and recommender systems. 
\end{IEEEbiography}

\end{document}